\documentclass[oldversion]{aa}  
\usepackage{graphicx}
\usepackage{txfonts}
\usepackage{hyperref}
\usepackage{xcolor}
\newcommand{\hi}{{\sc H\,i}}
\newcommand{\HIbf}{\mbox{H\hspace{0.155 em}{\footnotesize \bf I}}}

\newcommand{\mhi}{$M_\mathrm{HI}$}

\newcommand{\mstar}{$M_\star$}
\newcommand{\Nhi}{$N$(\hi)}
\newcommand{\mJybeam}{mJy beam$^{-1}$}
\newcommand{\msun}{{M$_\odot$}}

\newcommand{\kms}{$\,$km$\,$s$^{-1}$}
\newcommand{\ltsima} {$\; \buildrel < \over \sim \;$}
\newcommand{\gtsima} {$\; \buildrel > \over \sim \;$}
\newcommand{\lta} {\lower.5ex\hbox{\ltsima}}
\newcommand{\gta} {\lower.5ex\hbox{\gtsima}}

\begin{document}

   \title{The MeerKAT Fornax Survey - I. Survey description and first evidence of ram pressure in the Fornax galaxy cluster}
   \titlerunning{The MeerKAT Fornax Survey - I}

   \author{
   P.~Serra \inst{1}\fnmsep\thanks{paolo.serra@inaf.it},
   F.~M.~Maccagni \inst{2,1},
   D.~Kleiner \inst{2,1},
   D.~Moln\'{a}r \inst{1},
   M.~Ramatsoku \inst{3,1},
   A.~Loni \inst{4,1},
   F.~Loi \inst{1},
   W.~J.~G.~de~Blok \inst{2,5,6},
   G.~L.~Bryan \inst{7},
   R.~J.~Dettmar \inst{8},
   B.~S.~Frank \inst{9,10,6},
   J.~H.~van~Gorkom \inst{7},
   F.~Govoni \inst{1},
   E.~Iodice \inst{11},
   G.~I.~G.~J\'{o}zsa \inst{12},
   P.~Kamphuis \inst{8},
   R.~Kraan-Korteweg \inst{6},
   S.~I.~Loubser \inst{13},
   M.~Murgia \inst{1},
   T.~A.~Oosterloo \inst{2,5},
   R.~Peletier \inst{5},
   D.~J.~Pisano \inst{6},
   M.~W.~L.~Smith \inst{14},
   S.~C.~Trager \inst{5},
   \and
   M.~A.~W.~Verheijen \inst{5}
          }
          
   \authorrunning{Serra et al.}

   \institute{
   INAF - Osservatorio Astronomico di Cagliari, Via della Scienza 5, I-09047 Selargius (CA), Italy
   \and
   Netherlands Institute for Radio Astronomy (ASTRON), Oude Hoogeveensedijk 4, 7991 PD Dwingeloo, the Netherlands
   \and
   Department of Physics and Electronics, Rhodes University, PO Box 94, Makhanda 6140, South Africa
   \and
   Armagh Observatory and Planetarium, College Hill, Armagh BT61 9DG, UK
   \and
   Kapteyn Astronomical Institute, University of Groningen, PO Box 800, NL-9700 AV Groningen, the Netherlands
   \and
   Department of Astronomy, University of Cape Town, Private Bag X3, Rondebosch 7701, South Africa
   \and
   Department of Astronomy, Columbia University, New York, NY 10027, USA
   \and
   Ruhr University Bochum, Faculty of Physics and Astronomy,  Astronomical Institute, 44780 Bochum, Germany
   \and
   South African Radio Astronomy Observatory, 2 Fir Street, Black River Park, Observatory, Cape Town, 7925, South Africa
   \and
   The Inter-University Institute for Data Intensive Astronomy, Department of Astronomy, University of Cape Town, Private Bag X3, Rondebosch, 7701, South Africa
   \and
   INAF - Astronomical Observatory of Capodimonte, Salita Moiariello 16, 80131, Naples, Italy
   \and
   Argelander-Institut f\"{u}r Astronomie, Auf dem H\"{u}gel 71, D-53121 Bonn, Germany
   \and
   Centre for Space Research, North-West University, Potchefstroom 2520, South Africa
   \and
   School of Physics and Astronomy, Cardiff University, Queens Buildings, The Parade, Cardiff CF24 3AA, UK
   \\
             }

   \date{Received February 2, 2023; accepted February 23, 2023}

  \abstract
  {The MeerKAT Fornax Survey maps the distribution and kinematics of atomic neutral hydrogen gas (\hi) in the nearby Fornax galaxy cluster using the MeerKAT telescope. The 12 deg$^2$ survey footprint covers the central region of the cluster out to $\sim R_\mathrm{vir}$ and stretches out to $\sim2 R_\mathrm{vir}$ towards south west to include the NGC~1316 galaxy group. The \hi\ column density sensitivity ($3\sigma$ over 25 \kms) ranges from $5\times10^{19}$ cm$^{-2}$ at a resolution of $\sim10''$ ($\sim1$ kpc at the 20 Mpc distance of Fornax) down to $\sim10^{18}$ cm$^{-2}$ at $\sim1'$ ($\sim6$ kpc), and slightly below this level at the lowest resolution of $\sim100''$ ($\sim10$ kpc). The \hi\ mass sensitivity ($3\sigma$ over 50 \kms) is $6\times10^5$ \msun. The \hi\ velocity resolution is 1.4 \kms. In this paper we describe the survey design and \hi\ data processing, and we present a sample of six galaxies with long, one-sided, star-less \hi\ tails (of which only one was previously known) radially oriented within the cluster and with measurable internal velocity gradients. We argue that the joint properties of the \hi\ tails represent the first unambiguous evidence of ram pressure shaping the distribution of \hi\ in the Fornax cluster. The disturbed optical morphology of all host galaxies supports the idea that the tails consist of \hi\ initially pulled out of the galaxies' stellar body by tidal forces. Ram pressure was then able to further displace the weakly bound \hi\ and give the tails their present direction, length and velocity gradient.}
     
   \keywords{
   galaxies: clusters: individual: Fornax -- galaxies: evolution --galaxies: interactions
               }

   \maketitle

\section{MeerKAT Fornax Survey background and goals}
\label{sec:intro}

The MeerKAT Fornax Survey\footnote{https://sites.google.com/inaf.it/meerkatfornaxsurvey} is designed to map the distribution and kinematics of atomic neutral hydrogen gas within and in between galaxies in the Fornax cluster, delivering at the same time 1.4 GHz radio continuum images (both total- and polarised intensity) and rotation-measure cubes of the cluster. The broad goal of the survey is to progress our understanding of how galaxies evolve in low-redshift, low-mass galaxy clusters, of which Fornax is the nearest example given its virial mass $M_\mathrm{vir} = 5\times 10^{13}$ \msun\ \citep{drinkwater2001b} and its 20 Mpc distance \citep{blakeslee2001,blakeslee2009,jensen2001,tonry2001}.

The  context of the MeerKAT Fornax Survey is our quest to understand the link between galaxy properties and their position in the cosmic web \citep[e.g.,][]{hubble1931,spitzer1951,oemler1974,dressler1980,larson1980,giovanelli1983,postman1984,cappellari2011b}. To first order the picture is relatively clear. Galaxies form, travel and evolve across a wide range of matter density in the cosmic web. This corresponds to a broad and time-dependent range of physical conditions in their environment, such as the temperature and density of the surrounding intergalactic medium (IGM), the number density of galaxies, and the motion of galaxies relative to one another and to the IGM. These conditions are a key driver of the mass and distribution of the cold interstellar medium (ISM) of galaxies through: \it i) \rm the hydrodynamical interaction between the IGM and galaxies' ISM and circumgalactic medium (CGM) \citep[e.g.,][]{gunn1972,cowie1977,nulsen1982}; \it ii) \rm the tidal interaction between the large-scale gravitational potential and galaxies \citep[e.g.,][]{bekki1999}; and \it iii) \rm the interaction of galaxies with one another \citep[e.g.,][]{gallagher1972,moore1996}. It is largely thanks to these processes that the flow of cold gas in and out of galaxies and, therefore, their star formation activity and optical appearance depend on their position in the cosmic web.

The details of this picture are difficult to understand because the interplay and balance between the above processes depend on a large number of variables. For example, the strength of the ram pressure exerted by the IGM on the ISM / CGM of galaxies is expected to increase with environment density because of the higher galaxy speed $v_\mathrm{gal}$ and IGM density $\rho_\mathrm{IGM}$ (\citealt{gunn1972} showed that ram pressure scales as $\rho_\mathrm{IGM} \cdot v^2_\mathrm{gal}$). Conversely, the effect of individual galaxy encounters is expected to decrease with increasing environment density because of the higher relative speed of the interacting galaxies (the energy of the interaction scales as $v^{-2}_\mathrm{gal}$; e.g., \citealt{mo2010}), which might partly offset the higher frequency of such encounters. Furthermore, all types of interactions affect galaxies in a way that depends on their mass (since the varying depth and shape of galaxies' gravitational potential determines their ability to hold on to their ISM / CGM; e.g., \citealt{mori2000,boselli2008}), on their exact location and orbit within a cluster \citep[e.g.,][]{vollmer2001,jaffe2018}, and on the local properties of the ICM \citep[e.g.,][]{kenney2004}.

In this context, low-mass clusters like Fornax are interesting because they are intermediate between massive clusters ($M_\mathrm{vir} \gtrsim 10^{14}$ \msun), where the importance of hydrodynamical effects such as ram-pressure stripping is observationally well established \citep[e.g.,][for recent reviews see \citealt{cortese2021} and \citealt{boselli2022}]{gavazzi1978,gavazzi1995,dickey1997,bravoalfaro2000,vollmer2001,vollmer2004,kenney2004,chung2007,chung2009,cortese2007,yoshida2008,scott2010,boselli2016,poggianti2019,ramatsoku2019,ramatsoku2020,deb2020}, and small groups ($M_\mathrm{vir} \lesssim 10^{13}$ \msun), where galaxy interactions are more relevant \citep[e.g.,][many such systems are included in \citealt{hibbard2001}; for a review see \citealt{cortese2021}]{yun1994,koribalski2003,english2010,serra2013,leewaddell2019}. The effect and balance between the various environmental processes is poorly constrained by observations for clusters like Fornax. This is exactly what the MeerKAT Fornax Survey is designed to address on the basis of deep, high-resolution radio data.

Our main observational target, the 21-cm emission line of atomic neutral hydrogen (hereafter, \hi), has historically been a prime tracer of environment-driven galaxy evolution. As recently reviewed by \cite{cortese2021} and \cite{boselli2022}, it has allowed astronomers to demonstrate that the lower star-formation activity of spiral galaxies in clusters is associated with a lower \hi\ mass (\mhi) compared to that of similar galaxies outside clusters \citep[e.g.,][]{sullivan1978,chamaraux1980,giovanelli1985}, and that this in turn follows from the truncation of the \hi\ disc \citep[e.g.,][]{warmels1988,cayatte1990}. Sensitive \hi\ imaging has also revealed several cases of \hi\ discs in the process of losing gas \citep[e.g.,][]{bravoalfaro2000,kenney2004,chung2007,ramatsoku2019,luber2022,molnar2022}. On-going wide-area \hi\ surveys are now starting to strengthen the statistical basis of these results, albeit at limited resolution and sensitivity \cite[e.g.,][]{adams2020,koribalski2020,wang2021}. The MeerKAT Fornax Survey follows this thread of \hi\ observations of galaxy clusters by performing a deep, high-resolution study of Fornax.


Fornax is an interesting target not only because it is the nearest low-mass cluster but also because of its on-going assembly. First, its brightest central galaxy NGC~1399 is involved in an interaction with the nearby, infalling early-type galaxy NGC~1404, causing the sloshing of the surrounding ICM  \citep{machacek2005,scharf2005,su2017,sheardown2018}. Furthermore, the spatial distribution of luminous early-type galaxies is highly asymmetrical: most of them are found in an elongated structure (just west of NGC~1399), which has not yet dispersed within the cluster. This is also the region hosting most of the intra-cluster light and most of the galaxies with low-surface-brightness features, asymmetric stellar halos, high fractions of accreted stellar mass and small metallicity gradients --- all results which imply recent / on-going interactions \citep{iodice2016,iodice2017a,iodice2019,spavone2020,spavone2022}. Finally, Fornax hosts a diverse population of recent infallers: first, bright gas-rich late-type galaxies with moderate \hi-, H$_2$- and star formation rate deficiency \citep{zabel2019,loni2021,morokumamatsui2022}, which show evidence of environment-induced distortions in both the stellar body \citep{raj2019} and the ISM distribution \citep{leewaddell2018,zabel2019}; and second, star forming dwarf galaxies at the outskirts of the cluster covering a range of optical morphologies \citep{drinkwater2001a}. On the latter point we note that the number of known Fornax dwarfs has recently grown enormously thanks to the optical imaging work of \cite{munoz2015}, \cite{venhola2018,venhola2019,venhola2022} and \cite{ordenesbriceno2018}, but our knowledge of the dynamics of these galaxies within the cluster has not progressed much compared to what was presented by \cite{drinkwater2001a} because of the lack of new redshifts (see \citealt{maddox2019} for the latest compilation).

Despite the evidence of on-going growth of Fornax, radio observations have so far revealed only one clear case of interactions affecting the distribution of \hi\ in a galaxy falling into the cluster. That is NGC~1427A, whose \hi\ tail discovered by \cite{leewaddell2018} implies that a tidal interaction must have shaped the stellar body of the galaxy, while still allowing for the possibility that hydrodynamical processes are at work at the galaxy's outskirts (see also \citealt{mastropietro2021}). The most sensitive, comprehensive \hi\ survey of Fornax to date, covering the cluster out to $\sim R_\mathrm{vir}$ with \mhi\ sensitivity $\sim2\times10^7$ \msun\ and resolution $\sim1' \times 1.5'$, did not find any other such cases and revealed only a modest \hi\ deficiency of recent infallers (\citealt{loni2021}; for previous, single-dish radio surveys of Fornax see \citealt{bureau1996,barnes1997,schroeder2001,waugh2002,waugh2006}). \cite{loni2021} further argued that the ISM properties of galaxies entering Fornax change on long timescales ($\gtrsim$ 1-2 Gyr) presumably because of the low mass of the cluster.

Beyond the virial radius of Fornax, the galaxy group centred on NGC~1316 (host of the radio source Fornax A) and located $\sim 1.5$ Mpc ($\sim 2 R_\mathrm{vir}$) south west of the cluster centre exhibits some indications of interactions between its galaxies as they fall towards Fornax \citep{schweizer1980,mackie1998,goudfrooij2001,iodice2017b,raj2020,kleiner2021}. In particular, the MeerKAT commissioning observations taken in preparation for the MeerKAT Fornax Survey demonstrated that NGC~1316 itself formed about 1-2 Gyr ago through a 10:1 merger between a lenticular and a Milky Way-like galaxy \citep{serra2019}, as previously hypothesised by \cite{lanz2010}. Together with subsequent intra-group interactions \citep{schweizer1980,iodice2017b}, this event populated the IGM with neutral and ionised gas \citep{kleiner2021} and was shown to be a possible channel for injecting magnetic fields into the medium \citep{loi2022}. This dynamic environment provides enough cold gas to trigger the activity of the black hole at the centre of NGC~1316, whose rapid flickering on timescales of 10's of Myr has produced the complex radio continuum morphology of Fornax~A \citep{maccagni2020} and has impacted the distribution and kinematics of gas within this galaxy's stellar body \citep{maccagni2021}.

The MeerKAT Fornax Survey targets both the cluster central region and the NGC~1316 group, improving by 1-2 orders of magnitude over the sensitivity and resolution of previous \hi\ observations. It allows us to make a complete census of the on-going environmental interactions in Fornax by delivering \hi\ cubes, images, velocity fields and velocity dispersion maps with angular resolution from $\sim10''$ ($\sim1$ kpc) to $\sim100''$ ($\sim10$ kpc), velocity resolution of $\sim1.4$ \kms\ and column density sensitivity between $\sim8\times10^{17}$ and $\sim5\times10^{19}$ cm$^{-2}$ depending on angular resolution. Furthermore, it pushes the study of the \hi\ content of Fornax members down to \mhi\ $\sim6\times10^5$ \msun\ across the entire cluster, allowing us to measure the slope of the low-mass end of the \hi\ mass function \citep[thought to be sensitive to environmental effects, see][]{verheijen2001,springob2005,zwaan2005,pisano2011,moorman2014,jones2016,busekool2021} and to estimate the \hi\ deficiency of galaxies with stellar mass down to \mstar\ $\sim10^6$ \msun.

In this paper we: describe the MeerKAT Fornax Survey design and observations (Sect. \ref{sec:obs}); discuss the \hi\ data processing in order to provide a reference for all future papers using our \hi\ cubes and images (Sect. \ref{sec:reduction}); present a first key result, i.e., the first unambiguous evidence of ram-pressure shaping the properties of \hi-rich galaxies in Fornax (Sect. \ref{sec:results}); and provide a summary (Sect. \ref{sec:summary}).

\section{Survey design and observations}
\label{sec:obs}



\begin{figure}
\includegraphics[width=9cm]{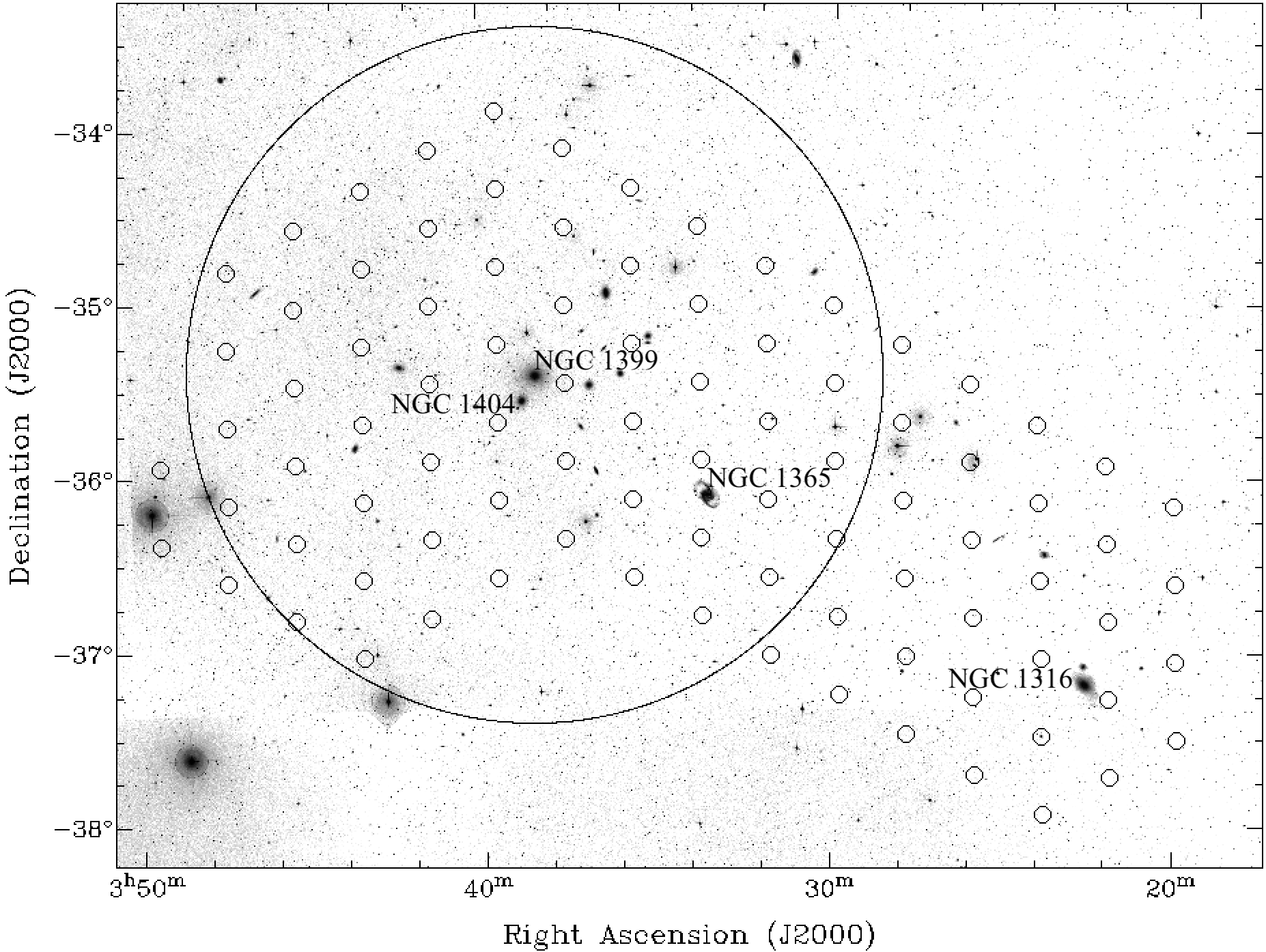}
\caption{MeerKAT Fornax Survey footprint (north is up, east is left). The small open circles overlaid on a Digitized Sky Survey red-band image represent the centre of the 91 MeerKAT pointings (Table \ref{table:radec}). Each pointing covers a fairly large field given the 1 deg FWHM of the MeerKAT primary beam at 1.4 GHz. The large open circle represents the virial radius $R_\mathrm{vir} \sim 700$ kpc. The brightest galaxies visible in the image are NGC~1399 at the cluster centre (with NGC~1404 just south east of it), the spiral NGC~1365 $\sim1$ deg south west of the cluster centre, and NGC~1316 (Fornax A) near the south west edge of the footprint.}
\label{fig:footprint}
\end{figure}

The MeerKAT Fornax Survey consists of L-band \hi\ and full-Stokes radio continuum imaging of a $\sim12$ deg$^2$ area centred on the Fornax galaxy cluster using MeerKAT \citep{camilo2018,mauch2020}. Figure \ref{fig:footprint} shows the survey footprint, which includes the full cluster out to approximately the virial radius $R_\mathrm{vir} \sim 700$ kpc \citep{drinkwater2001b} and extends to $\sim2 R_\mathrm{vir}$ towards south west, in the direction of the NGC~1316 galaxy group. We cover this area with 91 MeerKAT pointings distributed on a hexagonal mosaic grid with spacing 0.45 deg ($\sim1/2$ the MeerKAT primary beam FWHM at 1.4 GHz; \citealt{mauch2020}). Table \ref{table:radec} lists RA and Dec of all pointings. We integrate for 9 h on-source per pointing, which is designed to deliver a final \hi\ mosaic cube with a natural noise level of 0.1 \mJybeam\ when binning the data to 5-\kms-wide channels. This allows us to make sensitive \hi\ cubes and continuum images at a variety of resolutions (Sect. \ref{sec:mosaic}).

We observe each pointing of the MeerKAT Fornax Survey for $2\times5$ h including calibration overheads. In order to optimise the $uv$ coverage of each pointing we limit the hour angle overlap between the two 5-h observations: the first observation is done with Fornax rising, the second with Fornax setting. Table \ref{table:obs} lists the allowed LST range and the MeerKAT calibration and science scans of both observations. As described in Sec. \ref{sec:reduction}, we perform most data reduction steps independently on the \emph{rising} and \emph{setting} observation. We combine the two observations in the final stages of \hi\ imaging.

   \begin{table}
   {\centering
      \caption[]{Sequence of calibration and science scans of the \emph{rising} (top) and \emph{setting} (bottom) 5-h MeerKAT observations executed for each of the 91 pointings of the MeerKAT Fornax Survey (see Sect. \ref{sec:obs}). We use the primary calibrator for the time-independent bandpass, delay and flux-scale calibration. We use the secondary calibrator for the frequency-independent, time-dependent gain calibration. The polarisation calibrator is not used when reducing the 32k zoom data described in this paper. The total time on target for each pointing is 8.85 h.}
         \label{table:obs}
         \begin{tabular}{llll}
            \noalign{\smallskip}
            \hline
            \hline
            \noalign{\smallskip}
             repeat & type & field & integration \\
             &  &  & (min) \\
            \noalign{\smallskip}
            \hline
            \noalign{\smallskip}
             \multicolumn{4}{c}{\emph{(rising, 5 h integration, allowed LST range 22:30 - 05:00)}}\\
            \noalign{\smallskip}
            \hline
            \noalign{\smallskip}
            $1\times$ & primary cal & 1934-638 & 10  \\
            \noalign{\smallskip}
            \hline
            \noalign{\smallskip}
            $4\times$ & secondary cal & J0440-4333 & 2  \\
            & science & Fornax  & 53.6  \\
            \noalign{\smallskip}
            \hline
            \noalign{\smallskip}
            $1\times$ & polarisation cal & 3C138 & 5  \\ 
            & secondary cal & J0440-4333 & 2  \\
            & science & Fornax  & 53.6  \\
            & secondary cal & J0440-4333 & 2  \\
            & polarisation cal & 3C138 & 5  \\ 
            \noalign{\smallskip}
            \hline
            \noalign{\smallskip}
             \multicolumn{4}{c}{\emph{(setting, 5 h integration, allowed LST range 03:00 - 09:30)}}\\
            \noalign{\smallskip}
            \hline
            \noalign{\smallskip}
            $1 \times$ & primary cal & 0408-658 & 10 min \\
            \noalign{\smallskip}
            \hline
            \noalign{\smallskip}
            $5\times$ & polarisation cal & 3C138 & 3 min \\ 
            & secondary cal & J0440-4333 & 2 min \\
            & science & Fornax  & 52.6 min \\
            \noalign{\smallskip}
            \hline
            \noalign{\smallskip}
            $1\times$ & secondary cal & J0440-4333 & 2 min \\
            \noalign{\smallskip}
            \hline
            \noalign{\smallskip}
         \end{tabular}}
   \end{table}

We take the MeerKAT data simultaneously with two modes of the SKARAB correlator: \emph{i)} the 32k zoom mode for \hi\ science is tuned to the topocentric frequency range 1337-1444 MHz and delivers 32,768 channels, each with a width of 3.265 kHz; \emph{ii)} the 4k broad-band mode for radio continuum science samples the topocentric frequency range 856-1712 MHz with 4,096 channels, each with a width of 208.984 kHz. Both SKARAB modes deliver 4 linear correlations (HH,VV, HV, VH) with a dump time of 8 sec. In this paper we only describe the data obtained with the 32k zoom mode.

In order to maximise the sensitivity to diffuse \hi\ emission, our MeerKAT observations are executed when the fraction of available baselines meets the following constraints: $\geq80\%$ in the baseline length range [0,50] m; $\geq75\%$ in the ranges [50,100] m, [100,200] m, [200,400] m, [400,1000] m, [1,3] km; and $\geq50\%$ in the range [3,6] km. No constraints are given on baselines longer than 6 km or on the overall number of antennas. Furthermore, our observations are predominantly taken at night. The little data taken during daytime are typically affected by solar radio-frequency interference (RFI) as discussed below.

\section{\HIbf\ data processing}
\label{sec:reduction}

\subsection{Data transfer, computing resources and software}
\label{sec:idia}

We reduce the 32k zoom data on the Ilifu cloud facility\footnote{http://www.ilifu.ac.za}. We transfer only HH and VV correlations from the MeerKAT archive to Ilifu, and only channels 4,385 to 28,384 binned by a factor of 2. This channel selection excludes the bandpass roll-offs. The channel averaging results in 12,000 channels, each with a width of 6.531 kHz ($\sim1.4$ \kms\ for \hi\ at redshift $z=0$), covering the topocentric frequency range 1351-1428 MHz (approximate recessional velocity range from $-500$ to 14,700 \kms). Each 5-h raw dataset transferred to Ilifu in measurement-set (MS) format has a size of $\sim1.3$ TB, to which we add $\sim 3$ TB of temporary data during processing. Eventually, the fully calibrated and reduced MS ready for \hi\ imaging in the Fornax recessional velocity range has a size of $\sim0.5$ TB.

\begin{figure}
\includegraphics[width=9cm]{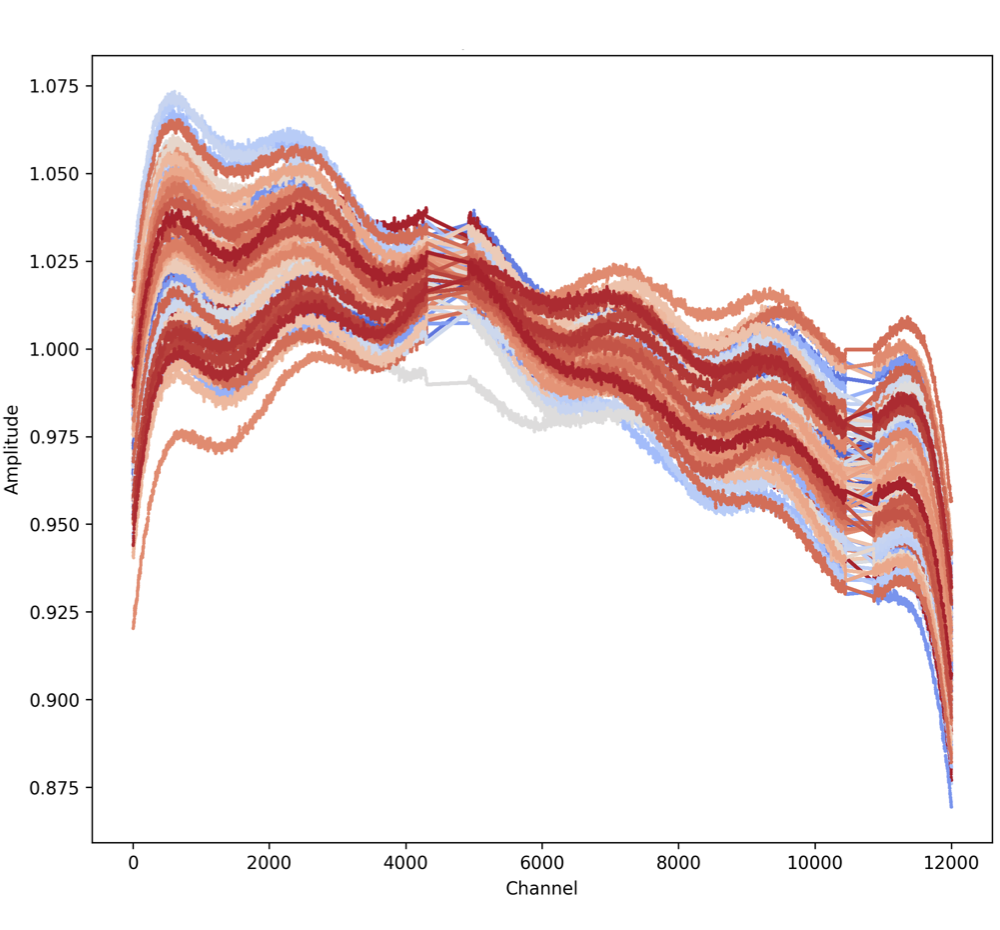}

\includegraphics[width=9cm]{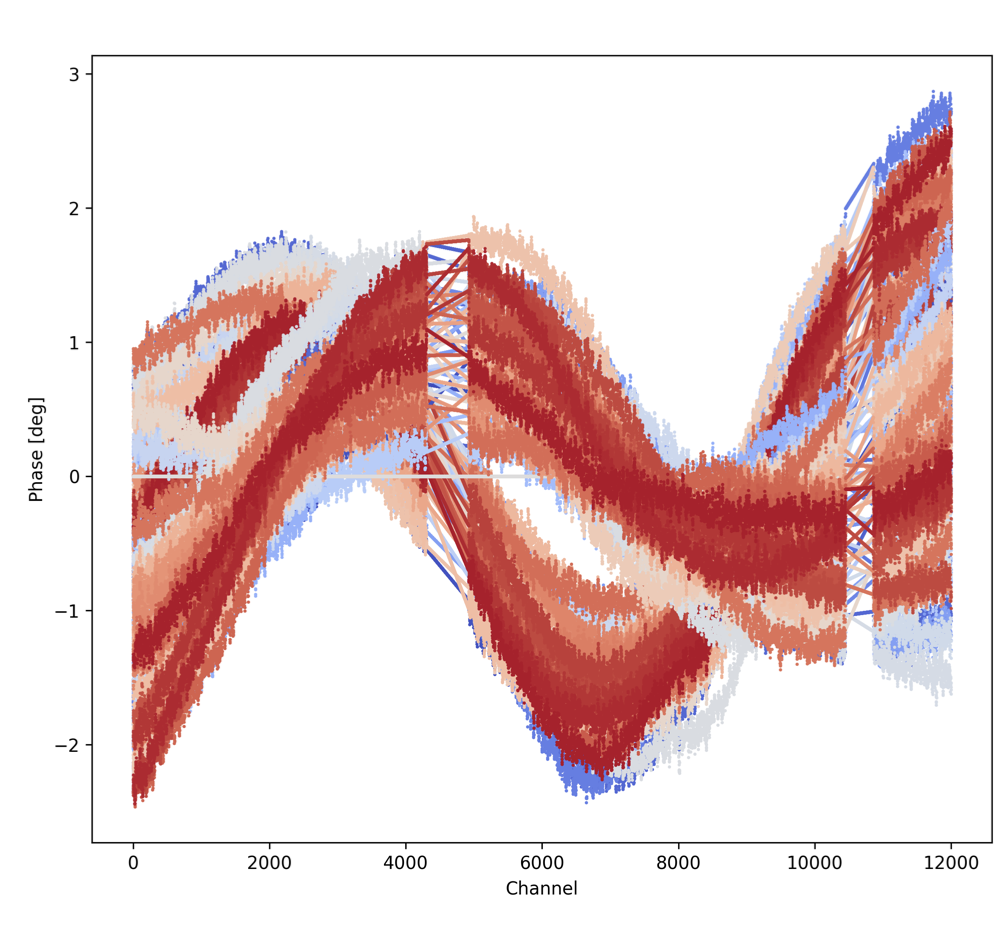}
\caption{Bandpass amplitude (top) and phase (bottom) of a typical MeerKAT 5-h observation. Each colour represents one of the 64 MeerKAT antennas. The bandpass is overall smooth and exhibits a high signal-to-noise ratio. The figure shows the result of interpolating linearly across the flagged frequency ranges.}
\label{fig:bp}
\end{figure}

\begin{figure}
\includegraphics[width=9cm]{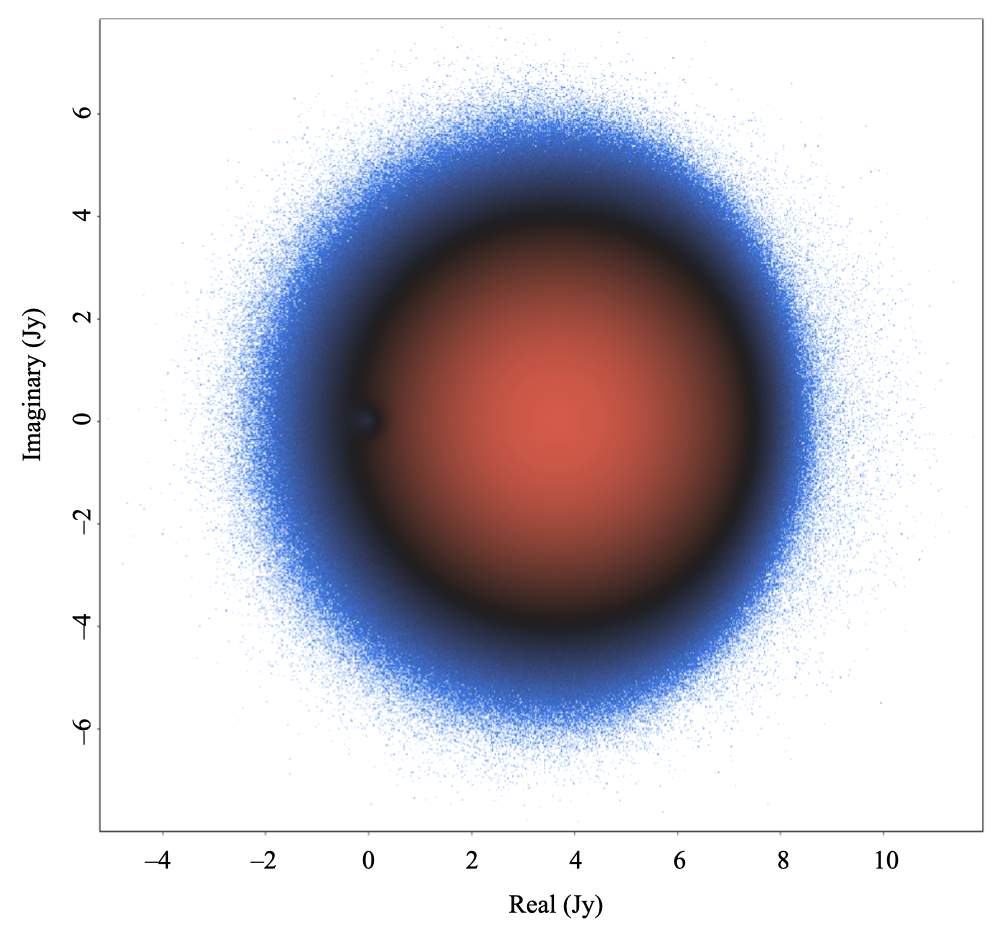}
\caption{Real-\emph{vs}-imaginary part of the HH cross-calibrated visibilities of the secondary calibrator for a typical MeerKAT 5-h observation. The colour represents the density of points on the plot. The distribution of points is perfectly circular in the inner region, where the density of points is higher, and shows some asymmetries only in the outer, sparsely-populated region. The standard deviation of the visibility amplitude and phase is typically $\sim$ 1 Jy and $\sim5$ deg, respectively.}
\label{fig:realimag}
\end{figure}

\begin{figure*}
\includegraphics[width=9cm]{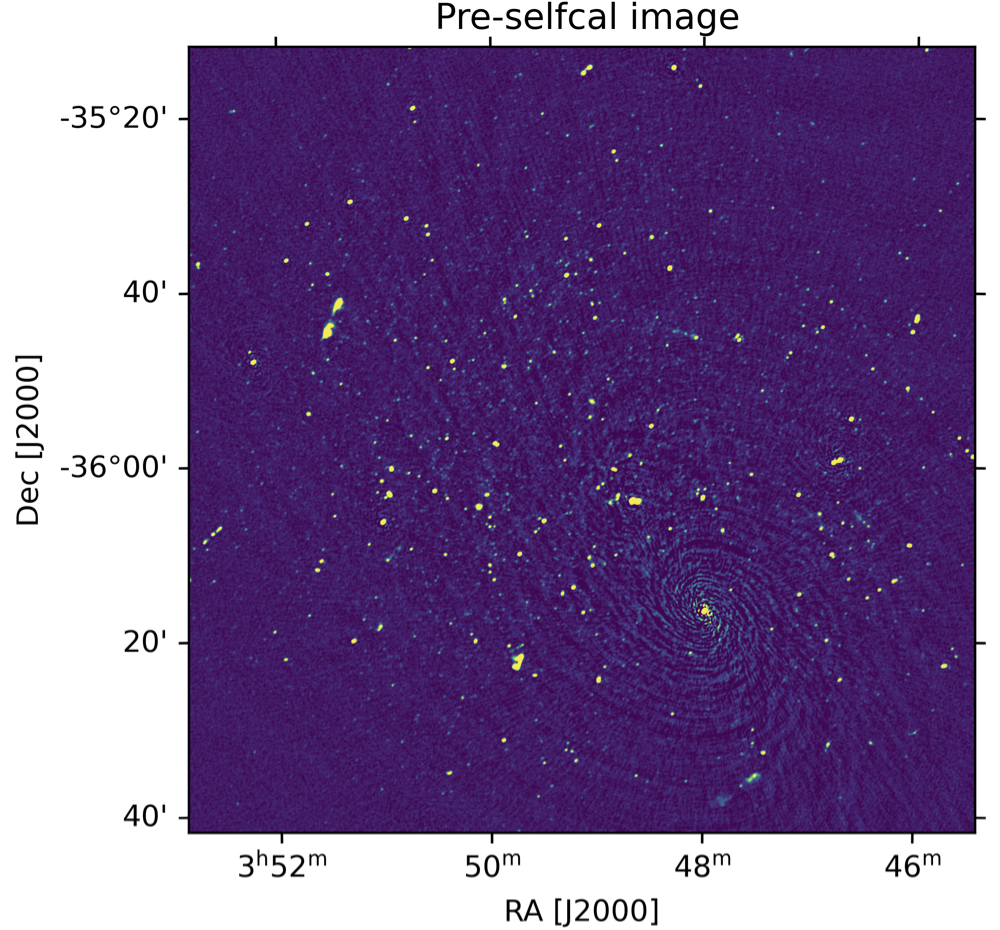}
\includegraphics[width=9cm]{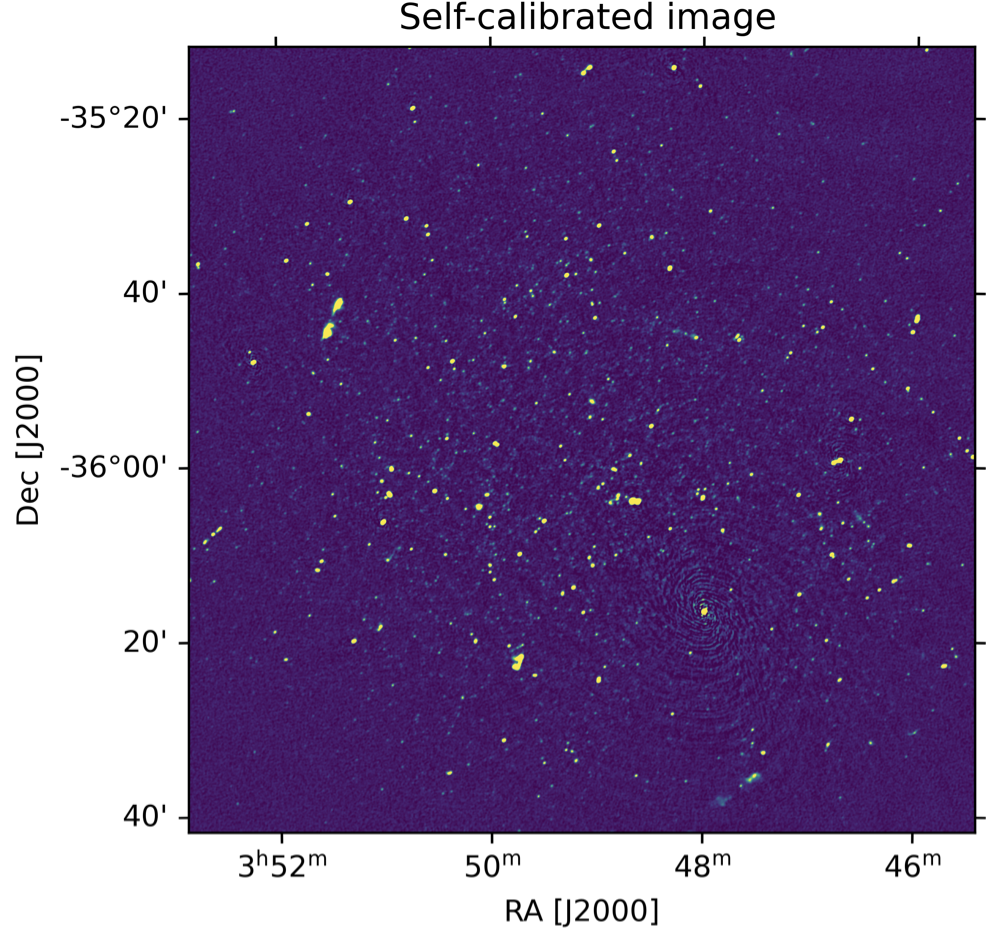}

\includegraphics[width=9cm]{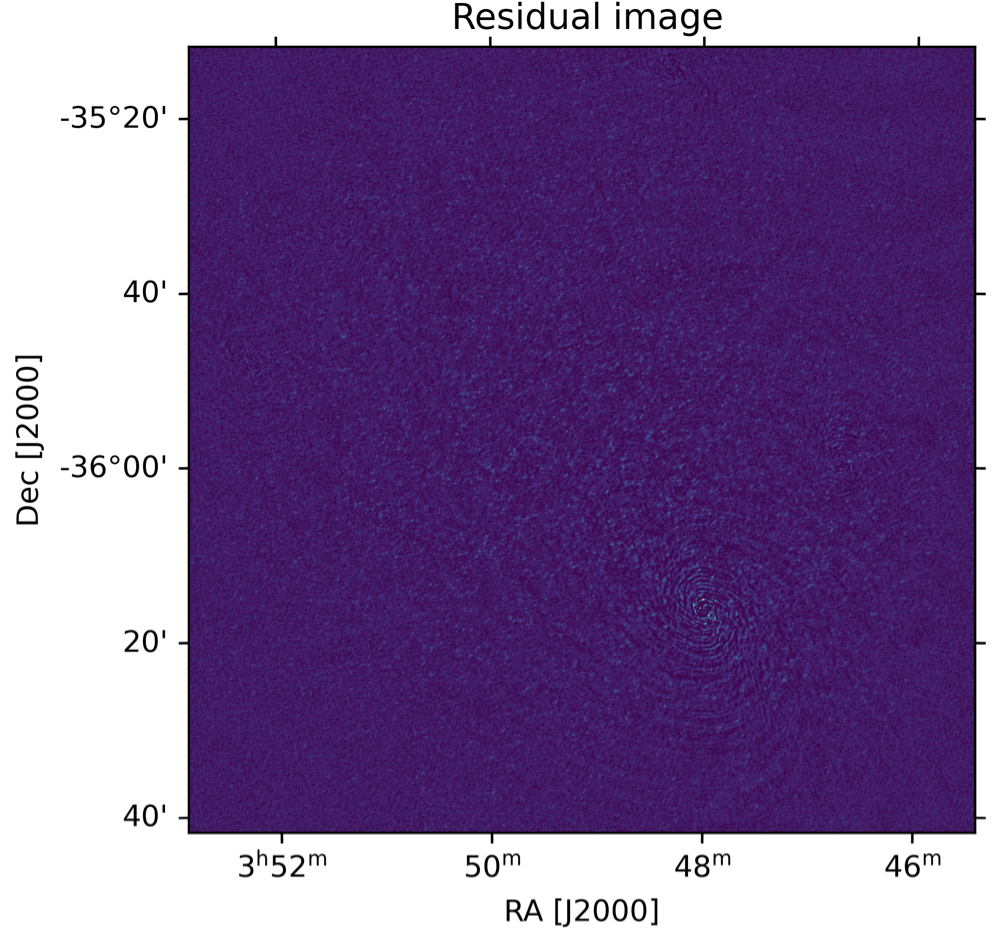}
\includegraphics[width=9cm]{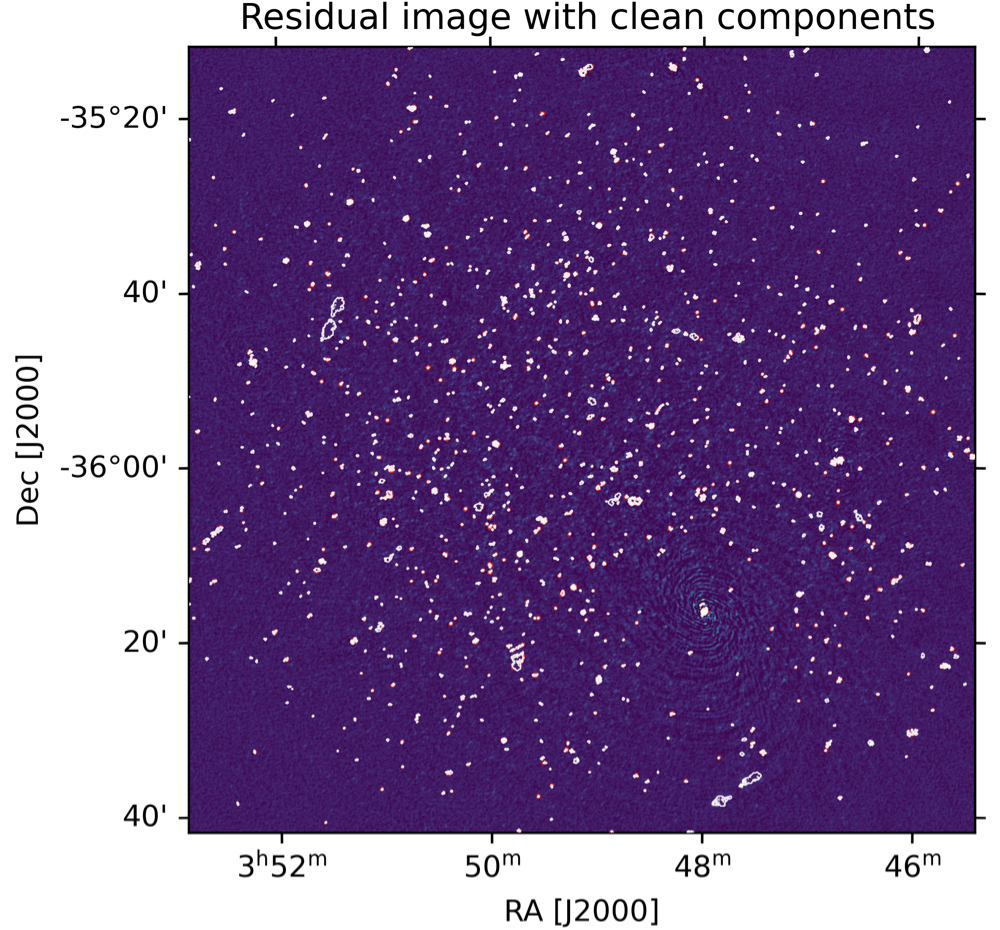}
\caption{Result of our imaging/self calibration loop for a typical MeerKAT 5-h observation. From left to right, top to bottom: the continuum image before self calibration; the continuum image after the last round of self calibration, showing the improved dynamic range; the last residual image, showing a population of faint sources not included in our sky model; and the residual image with white and red contours representing positive and negative clean components, respectively, demonstrating that our strategy minimises the inclusion of calibration artefacts around bright sources into the sky model. All images show a $2\times$ zoom-in of the full imaged field and are on the same linear colour scale from $-20$ to $+300$ $\mu$Jy beam$^{-1}$ (for comparison, the typical noise level is $\sim7$-10 $\mu$Jy beam$^{-1}$.}
\label{fig:cont}
\end{figure*}

We reduce the 32k zoom data of each 5-h observation using 4 Ilifu compute nodes. Each node has 32 CPUs and 230 GB of RAM. We reduce the data with the \texttt{CARACal} pipeline \citep{jozsa2020} as detailed in the rest of this Section. \texttt{CARACal} allows users to process their data employing a large number of radio interferometry software packages within a single pipeline thanks to \texttt{Stimela} \citep{makhathini2018}, a platform-independent radio interferometry scripting framework based on Python and container technology (in our case, \texttt{Singularity}; \citealt{kurtzer2017}). Below we specify which package we use for each step of our \texttt{CARACal} data reduction. When no package is given it means that we use tasks available as part of \texttt{CASA} \citep{mcmullin2007}. All data reduction steps described below are taken within \texttt{CARACal} unless stated otherwise. We validate the results of our data processing through visual inspection of the data products and diagnostic plots produced by \texttt{CARACal}, as illustrated throughout this Section.

\subsection{Calibrators flagging and cross calibration}
\label{sec:xcal}

We create an MS containing the primary and secondary calibrators (see Table \ref{table:obs}) and process it on a single Ilifu node. We flag autocorrelations, shadowed antennas and the frequency ranges 1379.6-1382.3 MHz (affected by the GPS L3 signal) and 1419.5-1421.3MHz (where emission/absorption from \hi\ in the Milky Way can corrupt the bandpass calibration). We then flag RFI with \texttt{AOFlagger} \citep{offringa2012} based on the Stokes Q visibilities, where the sky is faint and the RFI stands out. After all these steps, the typical calibrators' flagged fraction is $\sim10\%$.

We derive the cross calibration terms excluding baselines shorter than 100 m. For the the primary calibrator we use sky models that include confusing sources within the MeerKAT primary beam (see, e.g., \citealt{heywood2020}) and solve for antenna-based, time-independent delays, complex gains and complex bandpass in this order, repeating the sequence twice and applying at each step all calibration terms derived up to that point. We increase the bandpass signal-to-noise ratio by smoothing it with a 9-channel-wide mean running window (having verified that no genuine bandpass features exist on such scale). We plot the calibration solutions with \texttt{Ragavi}\footnote{https://github.com/ratt-ru/ragavi} for visual inspection. Fig. \ref{fig:bp} shows a typical bandpass solution.

We apply the primary calibrator's delay and bandpass to the secondary calibrator and solve for antenna-based, frequency-independent complex gains for every scan of the secondary independently (one solution per scan). We then flag the secondary's calibrated visibilities with \texttt{CASA}'s \texttt{tfcrop} to eliminate obvious outlying visibilities, and solve for the gains again. Finally, we scale the resulting gain amplitudes by bootstrapping the flux scale based on the primary calibrator's gains.

We apply the cross calibration terms to both the primary and secondary calibrator, and create a number of diagnostic plots with \texttt{Ragavi} for visual validation of the calibration (e.g., real-\emph{vs}-imaginary part of the calibrator's visibilities, amplitude-\emph{vs}-frequency and phase-\emph{vs}-frequency). We show a typical result in Fig. \ref{fig:realimag}. Altogether, the steps described in this subsection take typically $\sim4$ h using a single Ilifu node.

\subsection{Target flagging and cross calibration}
\label{sec:calflag}

We create four target MS files, each covering a $\sim20$ MHz sub-band equal to about 1/4 of the full band being processed. Each MS includes 3,000 unique channels plus 100 channels overlapping with the adjacent sub-band(s). We process these four MS files in parallel on four Ilifu nodes. We apply the primary calibrator's delays and bandpass as well as the secondary calibrator's gains on the fly while creating the four MS files. We flag autocorrelations and shadowed antennas, and then flag RFI with \texttt{AOFlagger} \citep{offringa2012} based on the Stokes Q visibilities using a slightly more aggressive flagging strategy than for the calibrators. We inspect the distribution of flags using images made with \texttt{RFInder}\footnote{https://github.com/Fil8/RFInder}, which show the flagged fraction as a function of frequency and baseline length for 10 minute-long time intervals. The typical flagged fraction is $\sim5\%$. We return to these MS files in Sect. \ref{sec:contsub}.

We bin the 3,000 unique channels of each MS by a factor of 150, creating four continuum MS files each with 20 0.980-MHz-wide channels. We use these MS files for continuum imaging and self calibration (Sect. \ref{sec:cont}). Altogether, the steps described in this subsection take $\sim6$ h on each of the four Ilifu nodes.

\subsection{Continuum imaging and self calibration}
\label{sec:cont}

Our next data processing step is to obtain a continuum model and self calibration adequate for continuum subtraction and \hi\ imaging. Continuum images for science analysis are obtained separately based on the 4k broad-band MeerKAT data and  will be described in a future paper.

\begin{figure}
\centering
\includegraphics[width=9cm]{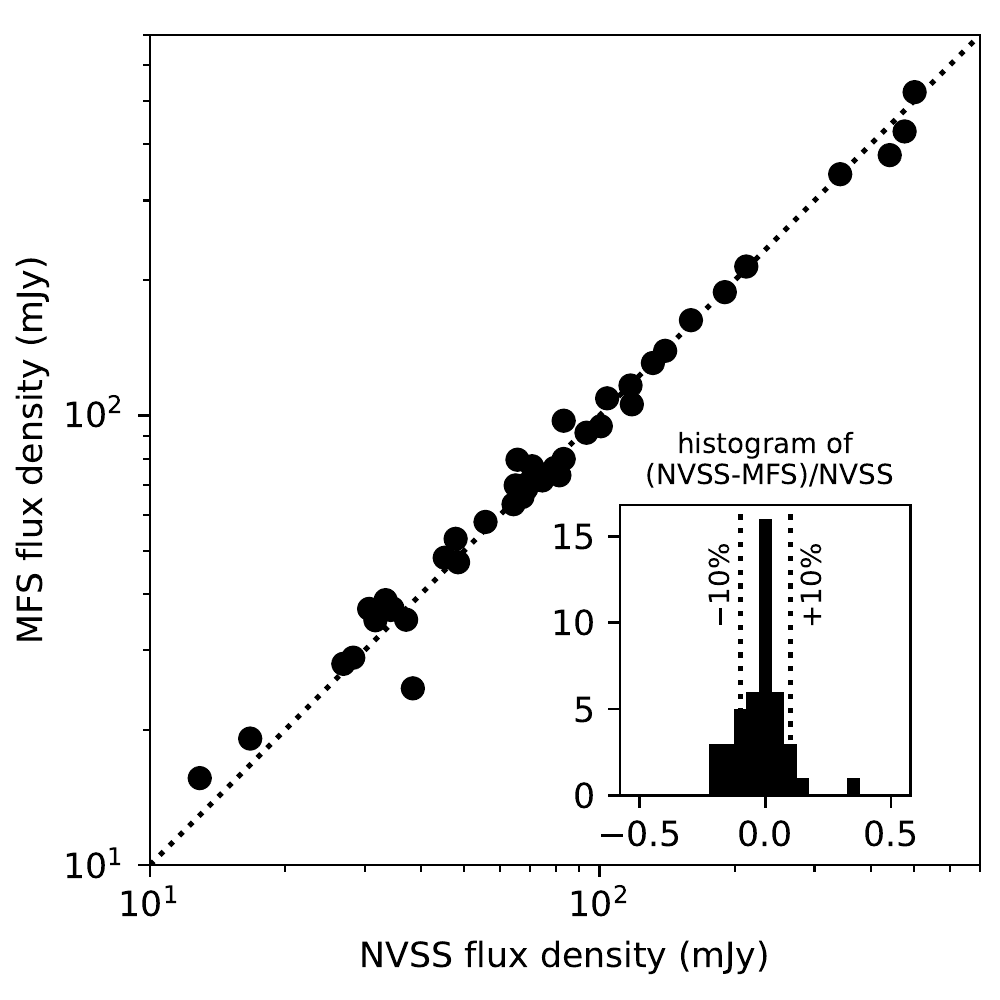}
\caption{Comparison between the MeerKAT Fornax Survey (MFS) and the NRAO VLA Sky Survey (NVSS) flux density of 44 radio continuum sources, showing the accuracy of our flux calibration. The dotted line in the main panel is the 1:1 relation. The inset shows the distribution of relative flux density differences, with the dotted lines marking the $\pm10\%$ levels. See text for details.}
\label{fig:fluxscale}
\end{figure}

For this part of the processing we use a single Ilifu node. We concatenate the four continuum MS files created in Sect. \ref{sec:calflag} into a single MS file with 80 0.980-MHz-wide channels. We flag the GPS L3 and Milky Way HI frequency ranges of the resulting MS as in Sect. \ref{sec:xcal}. Imaging is done with \texttt{WSclean} \citep{offringa2014,offringa2017} using a $12''$ $uv$-tapering, $2700\times2700$ $4''$ pixels (field of view $= 3\times3$ deg$^2$) and Briggs $robust = -0.5$. We image in 4 frequency sub-bands, regularising the spectral shape of each spatial clean component by fitting a second-order polynomial along the frequency axis. The clean details vary during the imaging/self calibration loop as described below. Self calibration is done with \texttt{Cubical} \citep{kenyon2018} solving for frequency-independent gain phase every 2 min. The imaging/self calibration loop consists of the following steps.

\begin{itemize}
\item First image. This image is obtained using the automated clean method of \texttt{WSclean}, where blind clean with a cutoff set by us to $8\times$ the local noise (evaluated in a 100-pixel-wide window) is followed by further cleaning of the blindly-cleaned pixels down to a threshold which we set to $0.5\times$ the local noise.
\item First clean mask. This is obtained running \texttt{SoFiA} \citep{serra2015a,westmeier2021} on the above cleaned image. We use the smooth+clip source finding algorithm applying Gaussian smoothing kernels with FWHM $0''$, $12''$, $24''$, $48''$, $96''$ and a detection threshold of $4.5\times$ the local noise evaluated as above.
\item Second image. This image is obtained cleaning within the above \texttt{SoFiA} clean mask down to $0.5\times$ the local noise.
\item First self calibration.
\item Second (deeper) clean mask. This is based on the last cleaned image, this time with a detection threshold of $4\times$ the local noise to include more sources in the clean model.
\item Third image. This image is obtained cleaning within the above \texttt{SoFiA} clean mask down to $0.5\times$ the local noise.
\item Second and last self calibration. These solutions will be applied to the \hi\ data (Sect. \ref{sec:contsub}).
\item Third and last (shallower) clean mask. This is based on the last cleaned image, this time with a detection threshold of $5\times$ the local noise. The purpose of this mask is to deliver a set of clean components for continuum subtraction (described in Sect. \ref{sec:contsub}). In this case, a shallower mask reduces the number of clean components and thus the processing time in subsequent steps (Sect. \ref{sec:contsub}). Faint continuum sources not included in this mask are subtracted in a different way as explained in Sect. \ref{sec:contsub}.
\item Fourth and last image. This is obtained cleaning within the above \texttt{SoFiA} clean mask down to $0.5\times$ the local noise.
\end{itemize}

A special case is that of fields which ``see'' the diffuse radio continuum lobes of Fornax~A \citep{ekers1983,fomalont1989,maccagni2020}. First, for such fields the automated strategy just described fails to fully include the radio lobes in the clean mask. We therefore make use of a \texttt{CARACal} mode described by \cite{maccagni2020}, in which a user clean mask (in our case, a clean mask tailored to Fornax A) is merged with the automated \texttt{SoFiA} masks obtained at each step of the imaging/self calibration loop. A second issue is that deconvolving Fornax A using delta-function clean components leaves substantial  artefacts in the final image. We therefore make use of multi-scale deconvolution (available in \texttt{WSclean}) selecting the scales up to $\sim3'$. The quality of the results obtained with this approach can be seen in \cite{serra2019}, \cite{maccagni2020} and \cite{kleiner2021}.

Fig. \ref{fig:cont} shows an example of the results of our imaging/self calibration loop. Such images are used to validate the quality of the self calibration and continuum modelling for all observations. Altogether, the steps described in this subsection take typically $\sim4$ h. The typical number of clean components for fields not including Fornax~A is 20,000, or $\sim0.3\%$ of all pixels in the image. The typical noise level is $\sim7$-10 $\mu$Jy beam$^{-1}$.

\begin{figure}
\includegraphics[width=9cm]{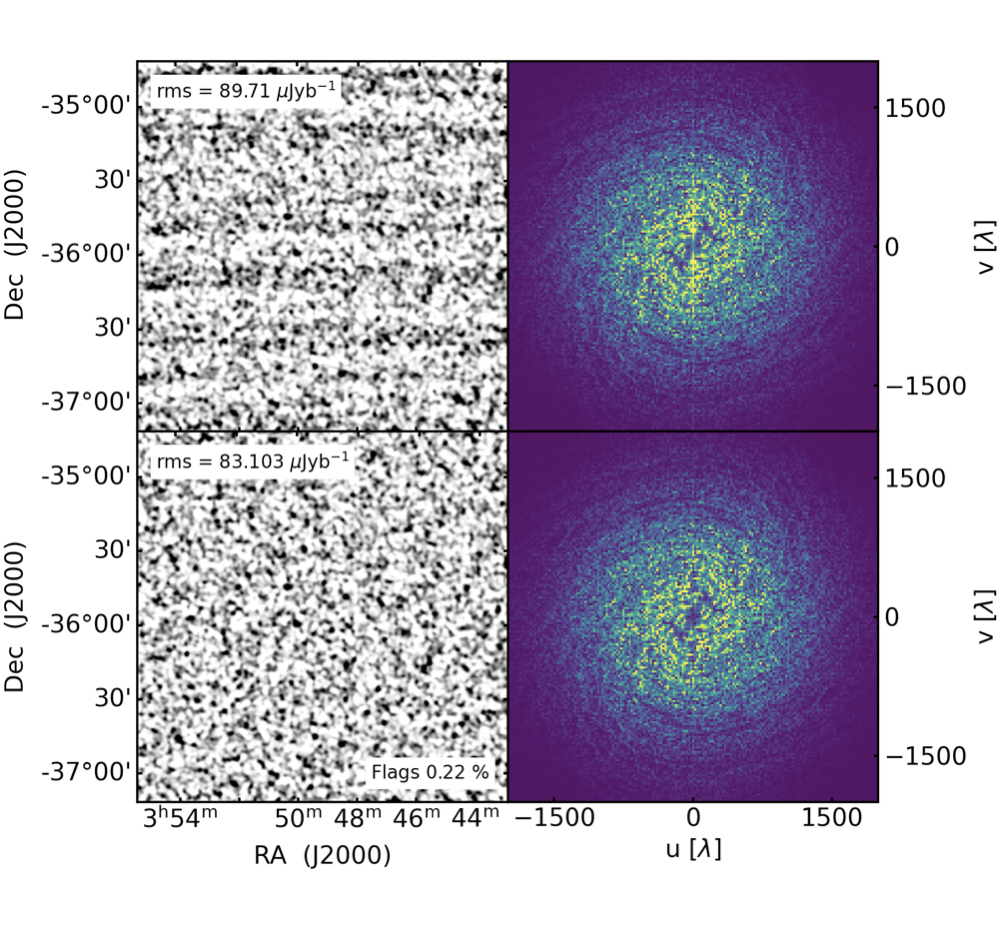}
\caption{\hi\ data before (top) and after (bottom) flagging the broad-band RFI near $u=0$ for a typical MeerKAT 5-h observation. The top row shows broad horizontal stripes in the image plane (left), and a corresponding bright vertical stripe at $u=0$ in its Fourier transform (right). The bottom row shows that, in this case, by flagging just 0.22\% of the data (as indicated in the bottom-right corner of the left panel) the stripes are no longer visible and the formal noise of the image decreases by 8\% (see noise values reported in the top-left corner of the left panels). Indeed, after flagging, the bright vertical stripe is no longer visible in the Fourier transform of the image (bottom-right panel). Note that this figure shows the image (and its Fourier transform) obtained combining all science scans of a 5-h observation, while the flags are calculated for each scan independently (see Sect. \ref{sec:uflag}).}
\label{fig:uflag}
\end{figure}

We measure the accuracy of our flux calibration by comparing the flux density of sources from the NRAO VLA Sky Survey (NVSS; \citealt{condon1998}) with the flux density measured from our radio continuum data. For this purpose we combine the images described in this subsection into a radio continuum mosaic using \texttt{MosaicQueen}\footnote{https://github.com/caracal-pipeline/MosaicQueen}. The frequency range covered by this mosaic (1351 - 1428 MHz) compares well with the two NVSS bands (approximately 1345 - 1385 MHz and 1415 - 1455 MHz, respectively), facilitating the flux density comparison. Figure \ref{fig:fluxscale} shows this comparison for a sample of 44 sources chosen to be bright and unresolved (or nearly so) in the NVSS. The sources are distributed uniformly within the area covered during the first two years of MeerKAT Fornax Survey observations (i.e., the eastern half of the footprint shown in Fig. \ref{fig:footprint}). The figure shows that our flux calibration is good. The distribution of relative flux density differences (inset in the figure) is centred on 0. The 16-th and 84-th percentiles are $-10.5\%$ and $+4.1\%$, respectively. Therefore, we estimate that the flux calibration contributes a $\pm10\%$ error to the \hi\ fluxes measured from our data.

\begin{figure}
\includegraphics[width=9cm]{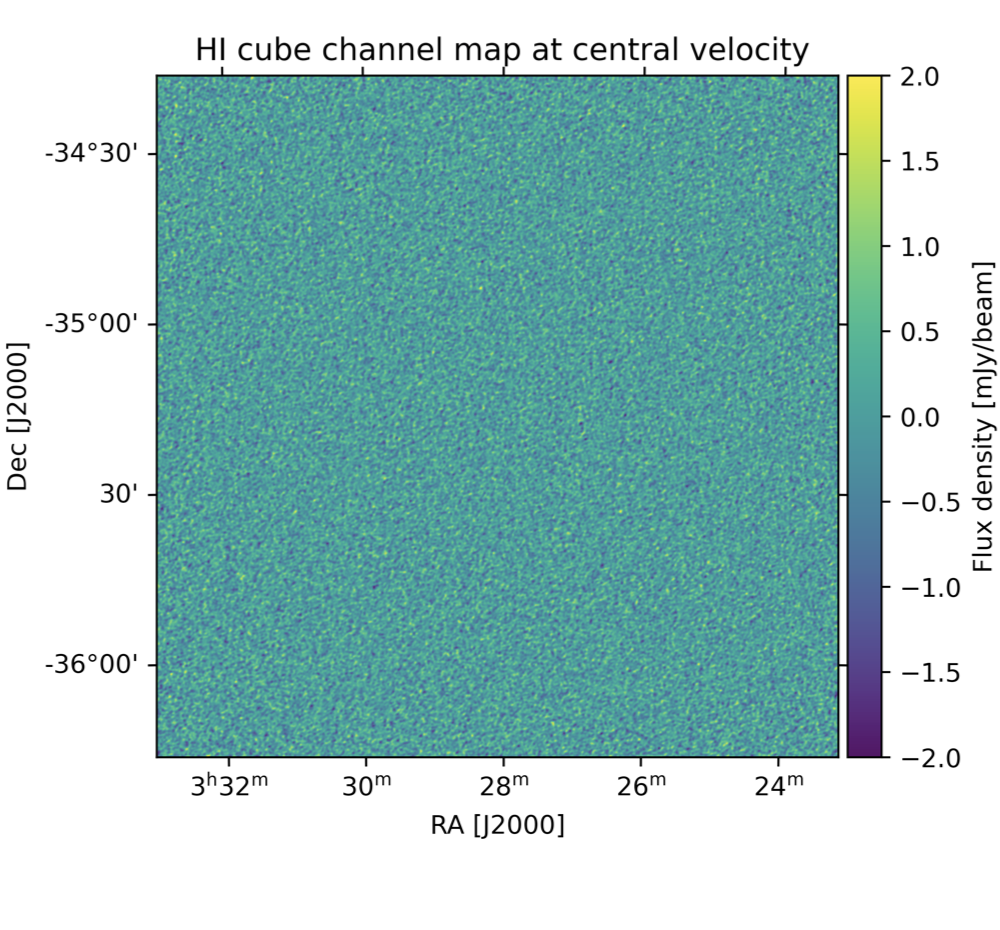}

\includegraphics[width=9cm]{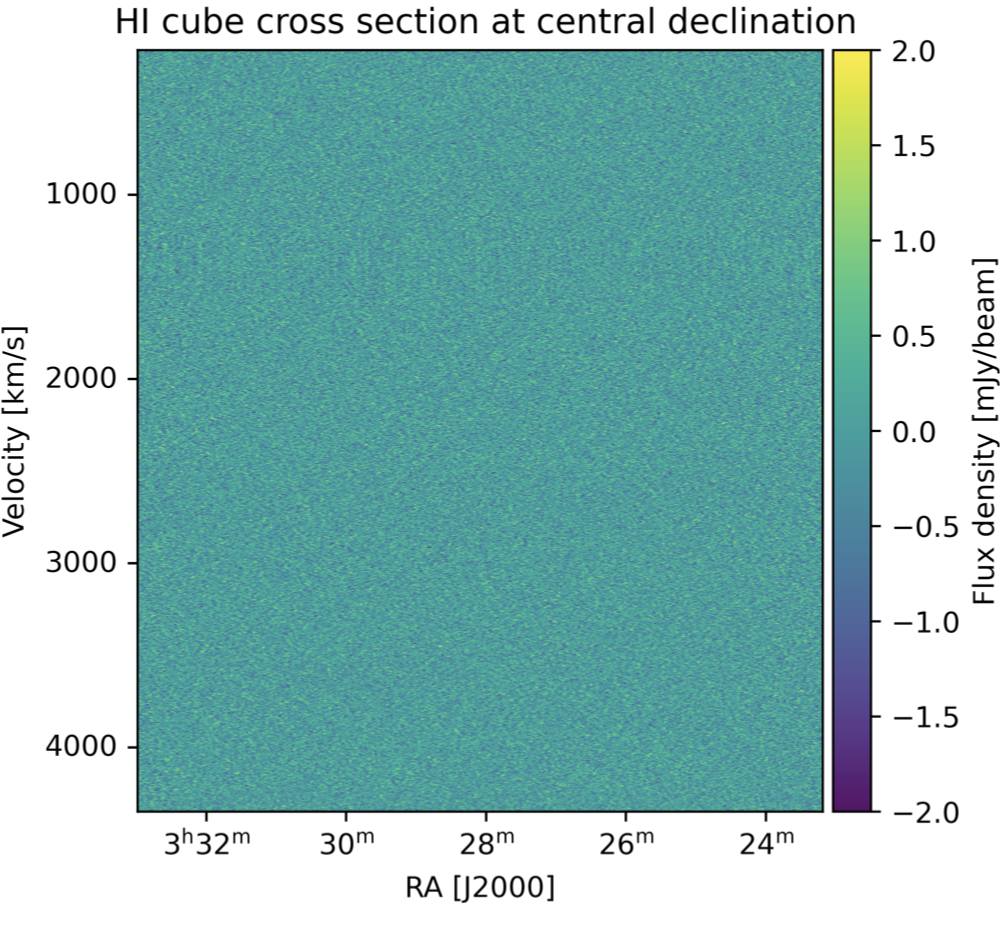}
\caption{Two slices through the $24''$-tapered \hi\ cube made for a typical MeerKAT 5-h observation. These slices show no artefacts due to residual RFI or continuum subtraction errors. Typically, the RA-velocity slices show some faint vertical structure in channel ranges excluded from the visibility spectral fit described in Sect. \ref{sec:contsub} (in this case, $\sim1200$ - $\sim1500$ \kms).}
\label{fig:xyz}
\end{figure}

\subsection{Continuum subtraction}
\label{sec:contsub}

For this part we use again four Ilifu nodes. Starting from the cross-calibrated MS files created in Sect. \ref{sec:calflag} we create four, narrower (in frequency) target MS files each covering about $\sim5$ MHz --- 1/16 of the full band being processed. Together, the four MS files cover generously the frequency range of interest for \hi\ in the Fornax cluster, from 1399.15 to 1420.04 MHz (100 - 4,500 \kms\ for the \hi\ line). We apply the self calibration gains to these four MS files with \texttt{Cubical}. We then Fourier-transform and copy the continuum clean components derived in Sect. \ref{sec:cont} to these MS files using \texttt{Crystalball}\footnote{https://github.com/caracal-pipeline/crystalball}. This is one of the most time-consuming steps of our data processing, and for this reason it is limited to the part of the band of interest for \hi\ in Fornax. We then subtract the continuum model from the target visibilities, thus removing most of the radio continuum emission from the data.

We refine the continuum subtraction by fitting and subtracting a first-order polynomial to each real and imaginary visibility spectrum independently, as in \cite{vanlangevelde1990}. Channels known to host \hi\ emission within 1 deg of the pointing centre are excluded from the fit. This step is done while simultaneously Doppler-correcting the data to a barycentric velocity grid, which we keep the same for all observations of the MeerKAT Fornax Survey. We thus obtain four, fully cross- and self-calibrated, continuum-subtracted, Doppler-corrected \hi\ MS files contiguous in barycentric velocity and ready for \hi\ imaging. Altogether, the steps described in this subsection take typically $\sim20$ h on each of the four Ilifu nodes.

\begin{figure}
\includegraphics[width=9cm]{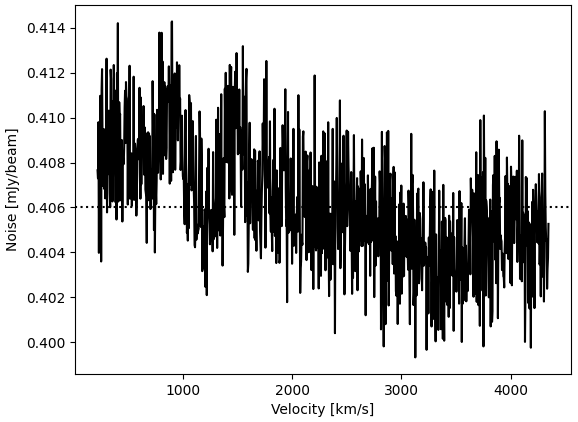}
\caption{Noise level as a function of \hi\ recessional velocity for the $24''$-tapered \hi\ cube made for a typical MeerKAT 5-h observation. The median noise level is indicated by the dashed line. As discussed in \cite{sault1994}, the noise increases in channels excluded from the visibility spectral fit described in Sect. \ref{sec:contsub} (in this case two small \hi\ velocity intervals centred at $\sim900$ and $\sim1300$ \kms, respectively).}
\label{fig:hinoise}
\end{figure}

\subsection{Flagging residual RFI}
\label{sec:uflag}

We continue processing the four \hi\ MS files on one Ilifu node each. Before \hi\ imaging we run \texttt{AOFlagger} with a shallow RFI-finding strategy to remove the rare but, in our experience, extremely bright artefacts that can be generated when fitting and subtracting the residual continuum emission from each visibility spectrum as described in Sect. \ref{sec:contsub}. These artefacts occur when all unflagged channels of a visibility spectrum are excluded from the fit because they contain \hi\ signal. In such rare cases (which should in principle produce an entirely flagged output spectrum), the unflagged channels excluded from the fit can have amplitude $\gg 1000$ Jy in the output spectrum, causing bright imaging artefacts.

We also run \texttt{Sunblocker}\footnote{https://github.com/gigjozsa/sunblocker} to flag any solar RFI present in our daytime data. Daytime data amount to just 1\% of all target visibilities of the MeerKAT Fornax Survey at the time of writing. In practice, the results of \texttt{Sunblocker} show that when solar RFI is noticeable in our \hi\ cubes we must flag all daytime visibilities with a baseline length below 1.5 k$\lambda$ ($\sim300$ m) in order to remove it from the data. This corresponds to angular scales $\gtrsim100''$ ($\sim10$ kpc at the distance of Fornax).

   \begin{table*}
   {\centering
      \caption[]{\hi\ mosaic cubes of the MeerKAT Fornax Survey}
         \label{table:mos}
         \begin{tabular}{llllllll}
            \noalign{\smallskip}
            \hline
            \hline
            \noalign{\smallskip}
             label & restoring beam$^\dagger$ & channel width & Briggs \emph{robust} & $uv$ taper & pixel size & noise & \Nhi$_{3\sigma, 25\mathrm{km/s}}$ \\
              & ($B_\mathrm{maj}\times B_\mathrm{min}$, $B_\mathrm{PA}$) & (\kms) &  &  &  & (\mJybeam) & (cm$^{-2}$) \\
            \noalign{\smallskip}
            \hline
            \noalign{\smallskip}
            $11''$ & $12.2''\times9.6''$, 142$^\circ$ & 1.4 & 0.0    & $6''$   & $2''$   & 0.30 & $5.0\times10^{19}$ \\ 
            $21''$ & $22.6''\times19.3''$, 135$^\circ$ & 1.4 & 0.0    & $15''$ & $5''$   & 0.26 & $1.2\times10^{19}$ \\ 
            $41''$ & $42.6''\times40.4''$, 115$^\circ$ & 1.4 & 0.5 & $30''$ & $10''$ & 0.24 & $2.7\times10^{18}$ \\ 
            $66''$ & $66.9''\times 65.0''$, 83$^\circ$ & 1.4 & 0.5 & $60''$ & $20''$ & 0.29 & $1.3\times10^{18}$ \\ 
            $98''$ & $100.2''\times95.8''$, 68$^\circ$ & 1.4 & 1.0 & $90''$ & $30''$ & 0.37 & $7.6\times10^{17}$ \\ 
            \noalign{\smallskip}
            \hline
            \noalign{\smallskip}
         \end{tabular}

         \small \emph{Notes.} ($\dagger$) Each channel of each field of the MeerKAT Fornax Survey has its own $uv$ coverage and, therefore, its own dirty beam. When cleaning with \texttt{WSclean} (Sect. \ref{sec:mosaic}) we set the restoring beam parameters to the fixed values given in this table. These are the median values of $B_\mathrm{maj}$, $B_\mathrm{min}$ and $B_\mathrm{PA}$ obtained as a fit to the main lobe of the dirty beam of each channel and each field during the first year of the survey (24/91 fields). Because of the homogeneity of our observation settings the distributions of these parameters are narrow. The 10$^\mathrm{th}$ and 90$^\mathrm{th}$ percentile levels of $B_\mathrm{maj}$ and $B_\mathrm{min}$ are within $\pm10\%$ of the median value; those of $B_\mathrm{PA}$ within $\pm10$ deg.\normalsize
         }
   \end{table*}

Finally, we run the flagging algorithm developed by \cite{maccagni2022} in order to remove the residual broad-band RFI typical of interferometers' short baselines near $u=0$ (e.g., \citealt{heald2016}). This RFI manifests itself as broad horizontal stripes in the \hi\ cubes and hampers accurate \hi\ source finding. As explained by \cite{maccagni2022}, an effective approach is to identify the \it uv \rm cells responsible for the RFI on the gridded \it uv \rm plane (Fourier-transform of the image plane), and flag all ungridded visibilities falling within those cells. This procedure can be applied to relatively long time intervals of the data independently --- in our case, each of the science scans listed in Table \ref{table:obs}. We calculate frequency-independent flags using data at 1350.8 - 1351.5 MHz, which is outside the Fornax velocity range and thus minimises contamination from \hi\ emission. We then apply the resulting flags to all channels given the broad-band nature of this RFI. We refer to \cite{maccagni2022} for details. Figure \ref{fig:uflag} shows an example of this flagging step. Altogether, the steps described in this subsection take typically $\lesssim 1$ h on each of the four Ilifu nodes, plus $\sim 1$ h on a single Ilifu node for flagging the RFI near $u=0$.

\subsection{Single-observation \hi\ imaging}
\label{sec:line}

We image the \hi\ using \texttt{WSclean}. When making these initial cubes, whose main use is data quality assurance, we bin channels by a factor of 3. This results in a cube with 250 channels with 4.1 \kms width for each of the four sub-bands formed in Sect. \ref{sec:contsub}. Imaging is done at two different resolutions, in both cases using Briggs $robust=0$ and a $2\times 2$ deg$^2$ field of view: \emph{i)} $900\times 900$ $8''$ pixels with a $24''$ $uv$-tapering; \emph{ii)} $360\times 360$ $20''$ pixels with a $60''$ $uv$-tapering. At both resolutions we clean the cubes blindly down to $6\times$ the noise level, after which the blindly-cleaned pixels are further cleaned down to $0.5\times$ the noise level (as in Sect. \ref{sec:cont} for the first continuum image). This round of cleaning does not include any major cycle (i.e., the Fourier transform and subtraction of clean components from the visibilities) since our goal is data quality assurance and the sidelobes of MeerKAT's dirty beam are very low. See Sect. \ref{sec:mosaic} for a description of the more accurate deconvolution performed on the final \hi\ cubes.

For each angular resolution we stack the cubes along the velocity axis in order to form a single cube, which covers the recessional velocity range 100 - 4,500 \kms. Fig. \ref{fig:xyz} shows two projections of a typical cube. We inspect all cubes visually in order to verify the absence of artefacts caused by, e.g., residual RFI or continuum subtraction errors.

We measure the noise level of these cubes as a function of \hi\ recessional velocity as shown in Fig. \ref{fig:hinoise}. At the time of writing, the noise level is within $\pm5\%$ of the expected value given the MeerKAT specs $T_\mathrm{sys}/\eta=20.5$ K at 1.4 GHz, where $T_\mathrm{sys}$ is the system temperature and $\eta$ is the aperture efficiency.

Altogether, the steps described in this subsection take typically $\sim3$ h on each of the four Ilifu nodes. They mark the end of the part of our data reduction procedure  which handles each 5-h MeerKAT observation separately. Below we describe how we combine the different observations in order to create \hi\ mosaic cubes suitable for scientific analysis.

\subsection{\hi\ mosaics}
\label{sec:mosaic}

The single-observation \hi\ cubes described in Sect. \ref{sec:line} were made for quality assurance binning channels by a factor of 3 and cleaning blindly with a relatively shallow threshold and no major cycles. In contrast, the \hi\ cubes for science analysis are made at the highest velocity resolution and with better cleaning. For each MeerKAT pointing we make new cubes by imaging the two 5-h observations together (\emph{rising} and \emph{setting}; see Table \ref{table:obs}) with no channel binning, resulting in a better dirty beam and in a channel width of 1.4 \kms. We make these cubes for the three sub-bands covering the Fornax recessional velocity range 200 - 3300 \kms\ starting from the lowest angular resolution in Table \ref{table:mos}, i.e., $98''$. We clean these cubes blindly using the source finder \texttt{SoFiA} within \texttt{CARACal} in order to automatically create clean masks and allowing \texttt{WSclean} to perform major cycles. During this process we adopt the same restoring beam for all channels, as detailed in Table \ref{table:mos}. We also create primary-beam cubes using the \cite{mauch2020} model. For each of the three sub-bands, we use \texttt{MosaicQueen} to linearly mosaic all pointings, truncating the primary beam at a response level of 10\%. We then stack the three resulting mosaic cubes along the velocity axis to form a single mosaic cube covering the full recessional velocity range of Fornax.

Having obtained the $98''$ \hi\ mosaic cube, we run \texttt{SoFiA} manually outside \texttt{CARACal} to create a better clean mask. We verify that this mask includes all (and only) real \hi\ emission detectable on various angular and velocity scales through visual inspection of 2D cube slices as well as in virtual reality (\texttt{iDaVIE}; \citealt{jarrett2021}). We use this clean mask to make our final $98''$ \hi\ cubes, which we then combine into our final $98''$ \hi\ mosaic. Subsequently, we use the same clean mask to image \hi\ at a resolution of $66''$ (Table \ref{table:mos}), combining the individual fields into the $66''$ \hi\ mosaic. We run \texttt{SoFiA} on the resulting cube as above to make a new clean mask, which we use to clean the $41''$ cubes --- and so on all the way to the highest resolution of $11''$. In fact, at the highest resolution we do not make a full-field \hi\ mosaic because it would be prohibitively large for visualisation and analysis. Instead, we make several small mosaic cubes centred on our \hi\ detections. The mask used for cleaning at a given resolution is also used to create \hi\ moment images and measure \hi\ properties (e.g., \hi\ fluxes) at that resolution.

Table \ref{table:mos} lists the parameters of the final \hi\ mosaic cubes made for the MeerKAT Fornax Survey\footnote{The table reports the specs of the mosaics at the time of writing, which cover approximately the eastern half of the survey footprint shown in Fig. \ref{fig:footprint}}. The noise levels listed in the table are the average values measured in the inner region of the mosaics, defined for this purpose as the region where the noise is $\leq\sqrt{2} \times$ the minimum noise level. The boundary of this area follows closely the distribution of the outer MeerKAT pointings in Fig. \ref{fig:footprint}. The noise level is constant within a few percent across most of this area, and increases rapidly further out (see also \citealt{serra2016b}). The \hi\ column density sensitivity given in the table is defined at a $3\sigma$ level and assuming a line width of 25 \kms. It varies from $\sim 5 \times10^{19}$ cm$^{-2}$ at $\sim11''$ resolution ($\sim 1.1$ kpc at the assumed Fornax distance of 20 Mpc) to $\sim 10^{18}$ cm$^{-2}$ at $\sim66''$ ($\sim 6.4$ kpc), reaching even lower column density at lower resolution. The $3\sigma$ \mhi\ sensitivity over 50 \kms\ is $\sim6\times10^5$ \msun\ at $20''$ - $40''$ resolution.

\section{First evidence of ram pressure in Fornax}
\label{sec:results}

\begin{figure*}
\centering
\includegraphics[width=9cm]{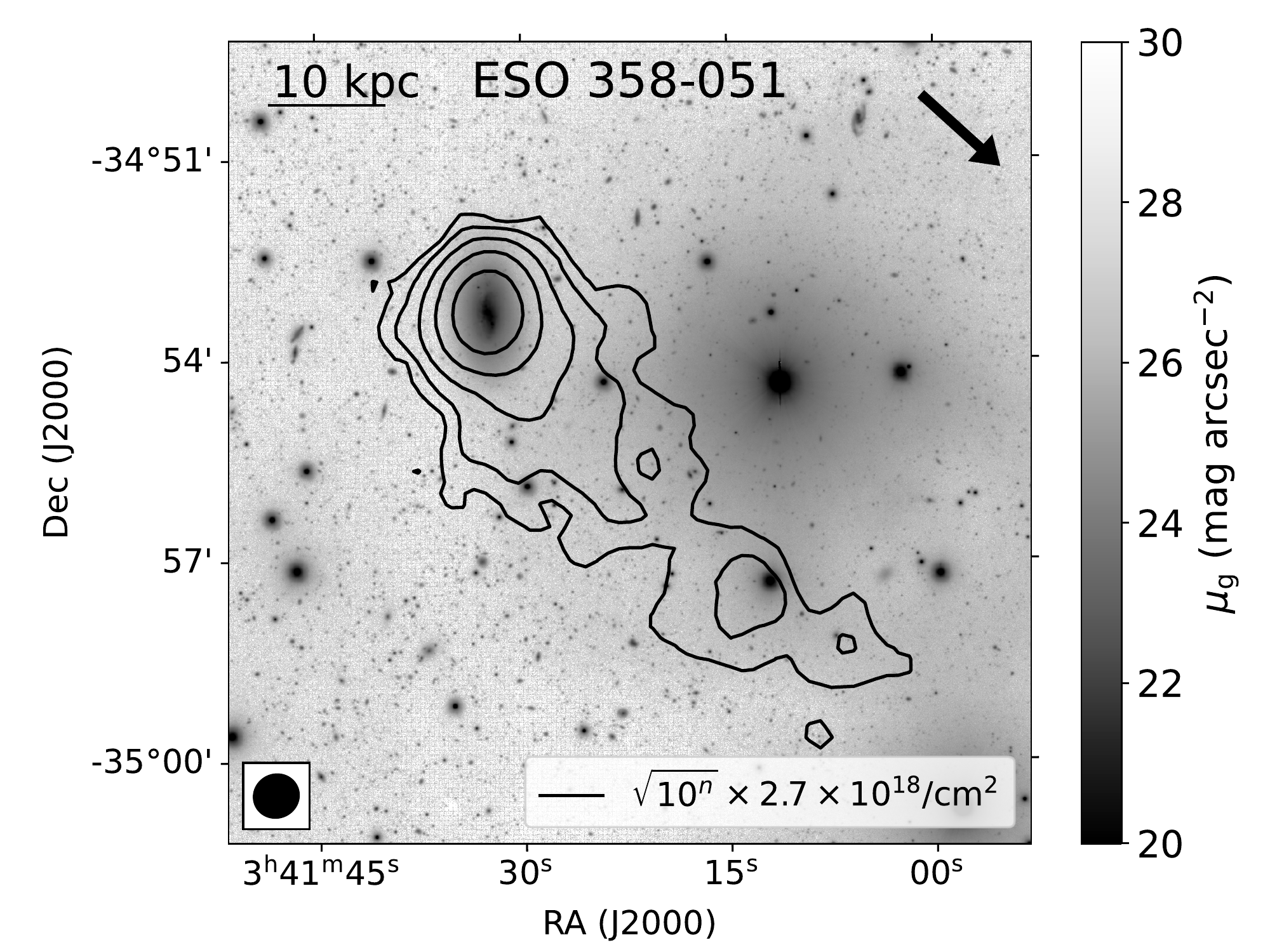}
\includegraphics[width=9cm]{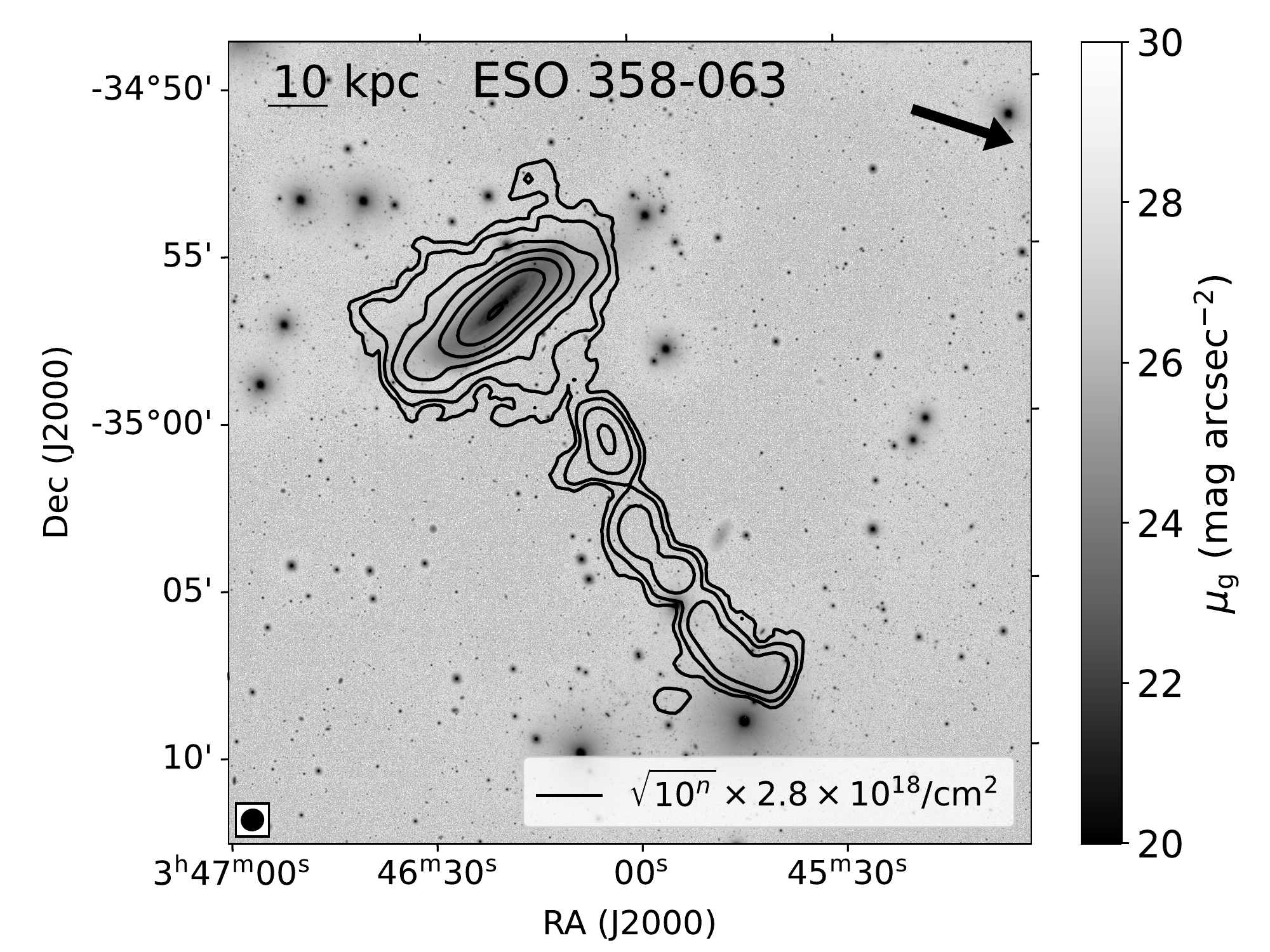}

\includegraphics[width=9cm]{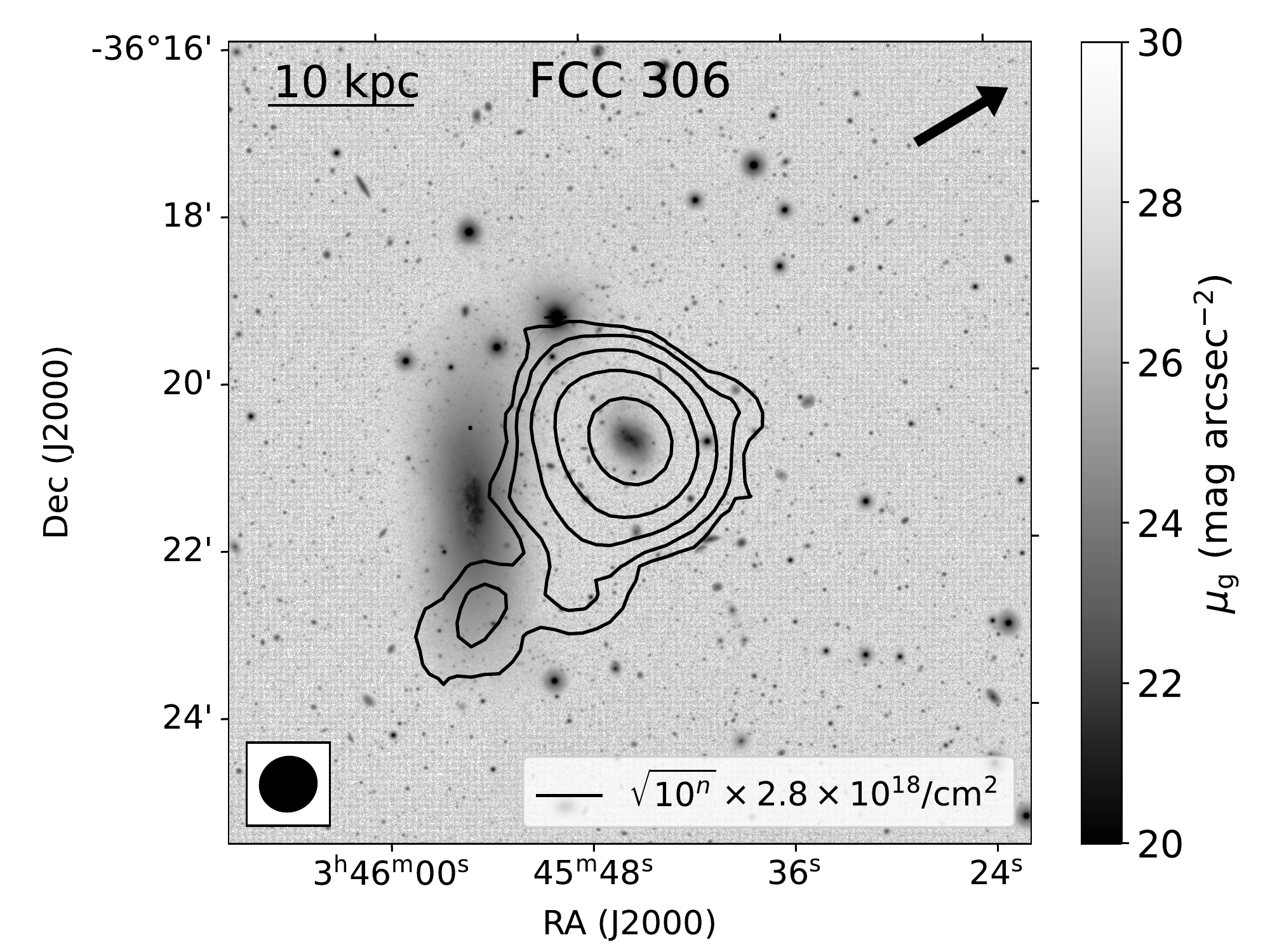}
\includegraphics[width=9cm]{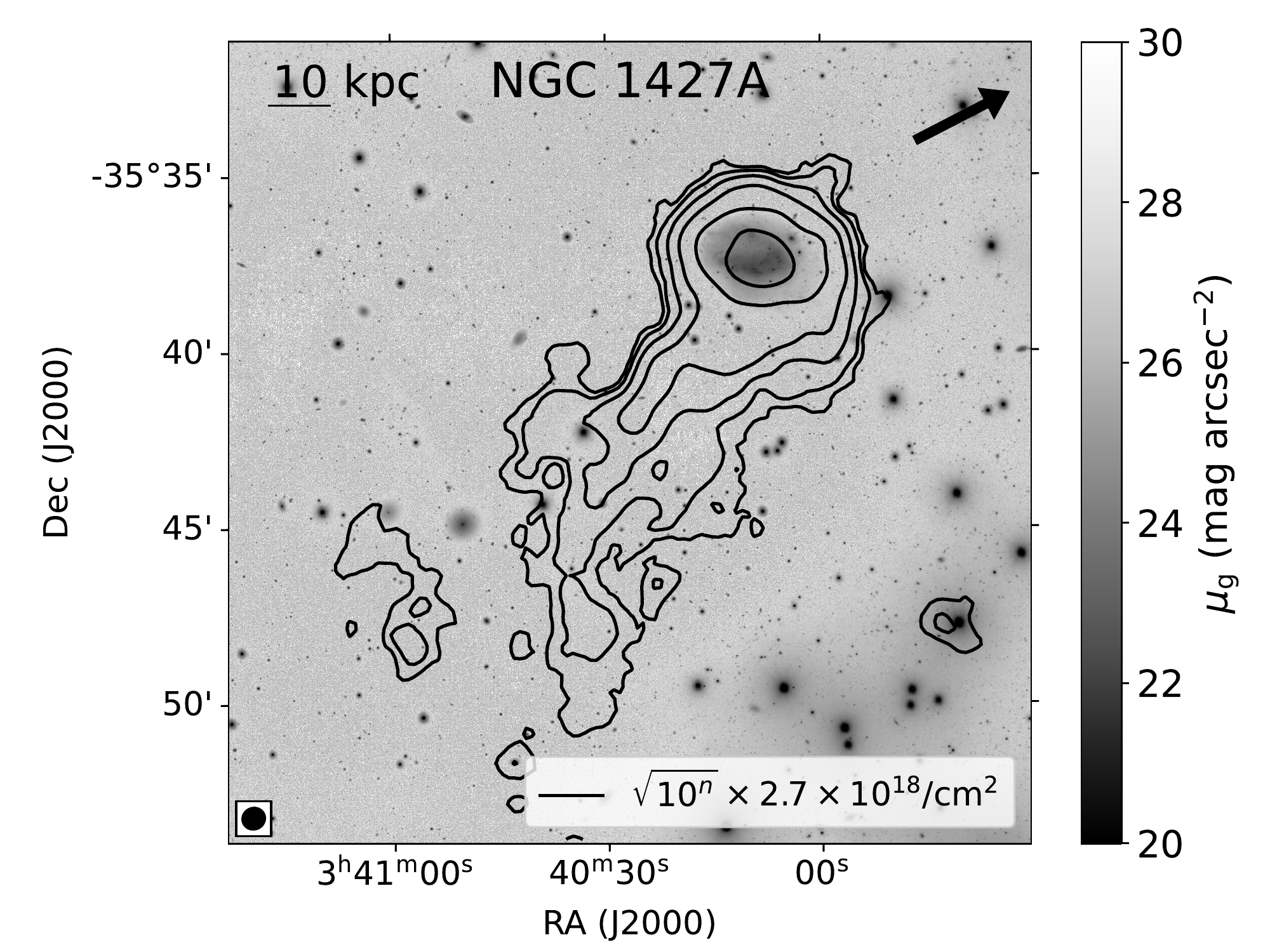}

\includegraphics[width=9cm]{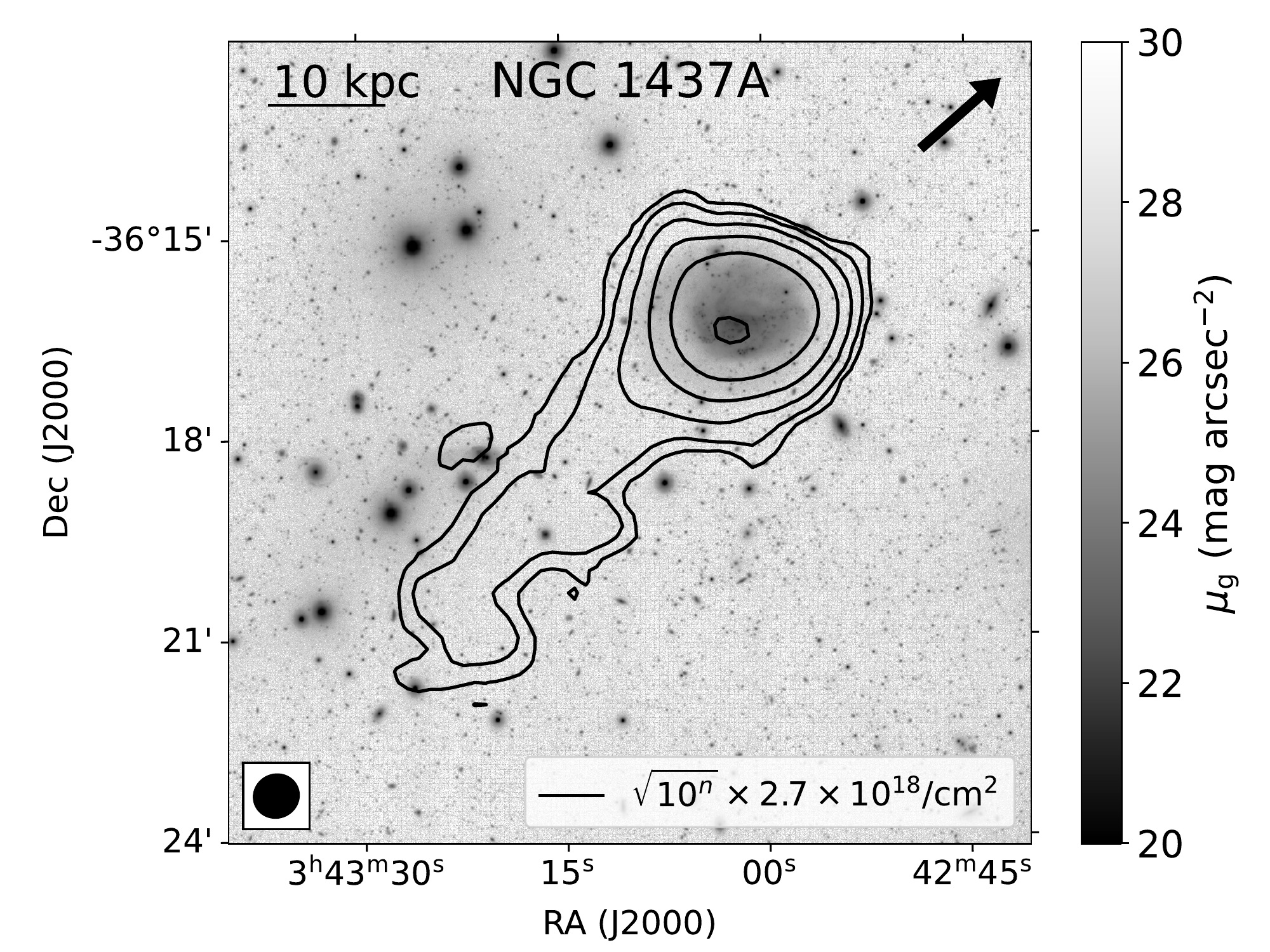}
\includegraphics[width=9cm]{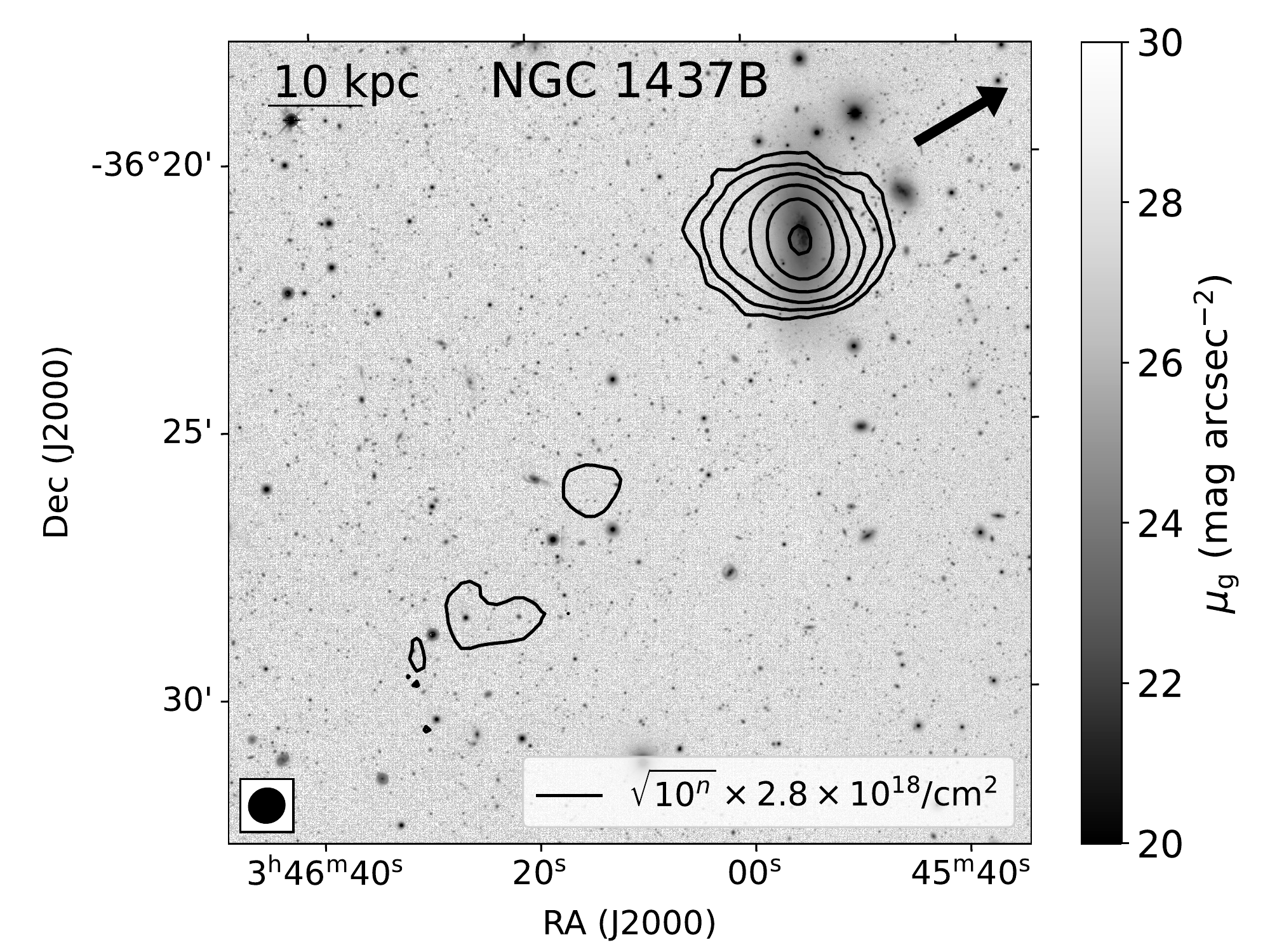}
\caption{MeerKAT \hi\ contours overlaid on a $g$-band image from the Fornax Deep Survey for the 6 galaxies with a one-sided \hi\ tail. The galaxy name is given at the top of each image. The \hi\ contour levels are given in the bottom-right legend ($n=0, 1, ...$). The bottom-left black ellipse represents the $41''$ \hi\ resolution. The scalebar in the top-left corner indicates 10 kpc. The top-right arrow points towards NGC~1399 at the centre of Fornax.}
\label{fig:tails}
\end{figure*}

\begin{figure*}
\centering
\includegraphics[width=8.5cm]{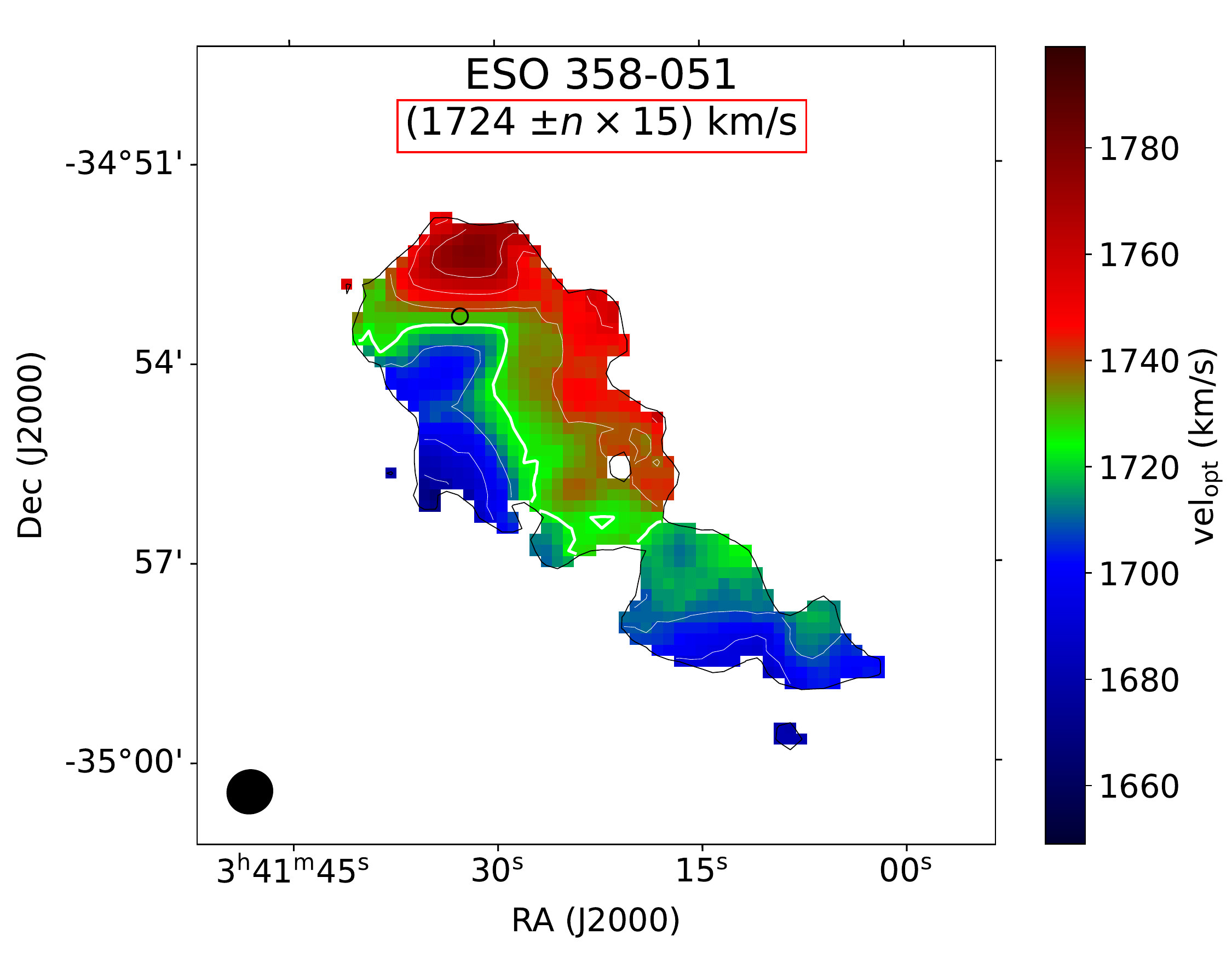}
\includegraphics[width=8.5cm]{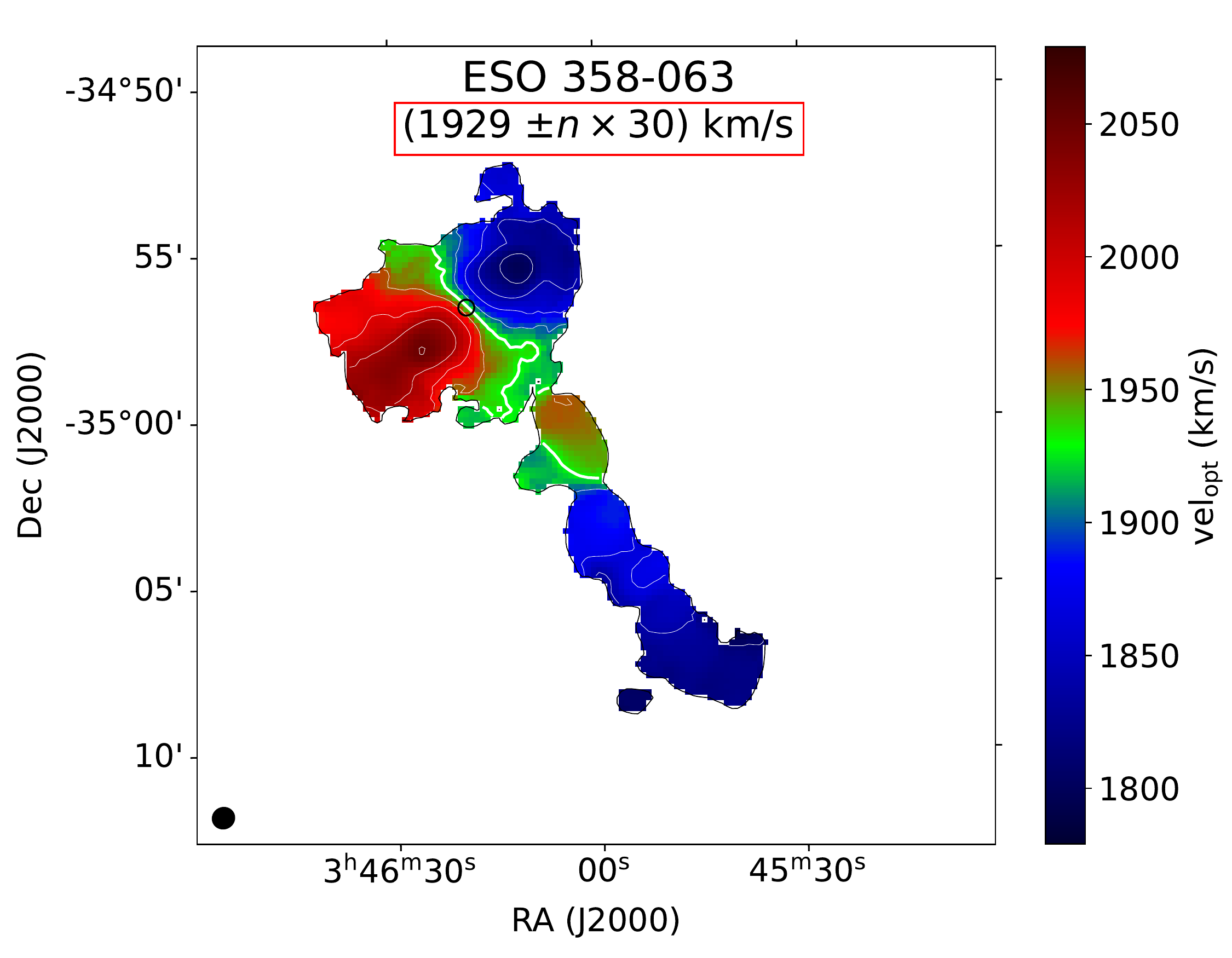}

\includegraphics[width=8.5cm]{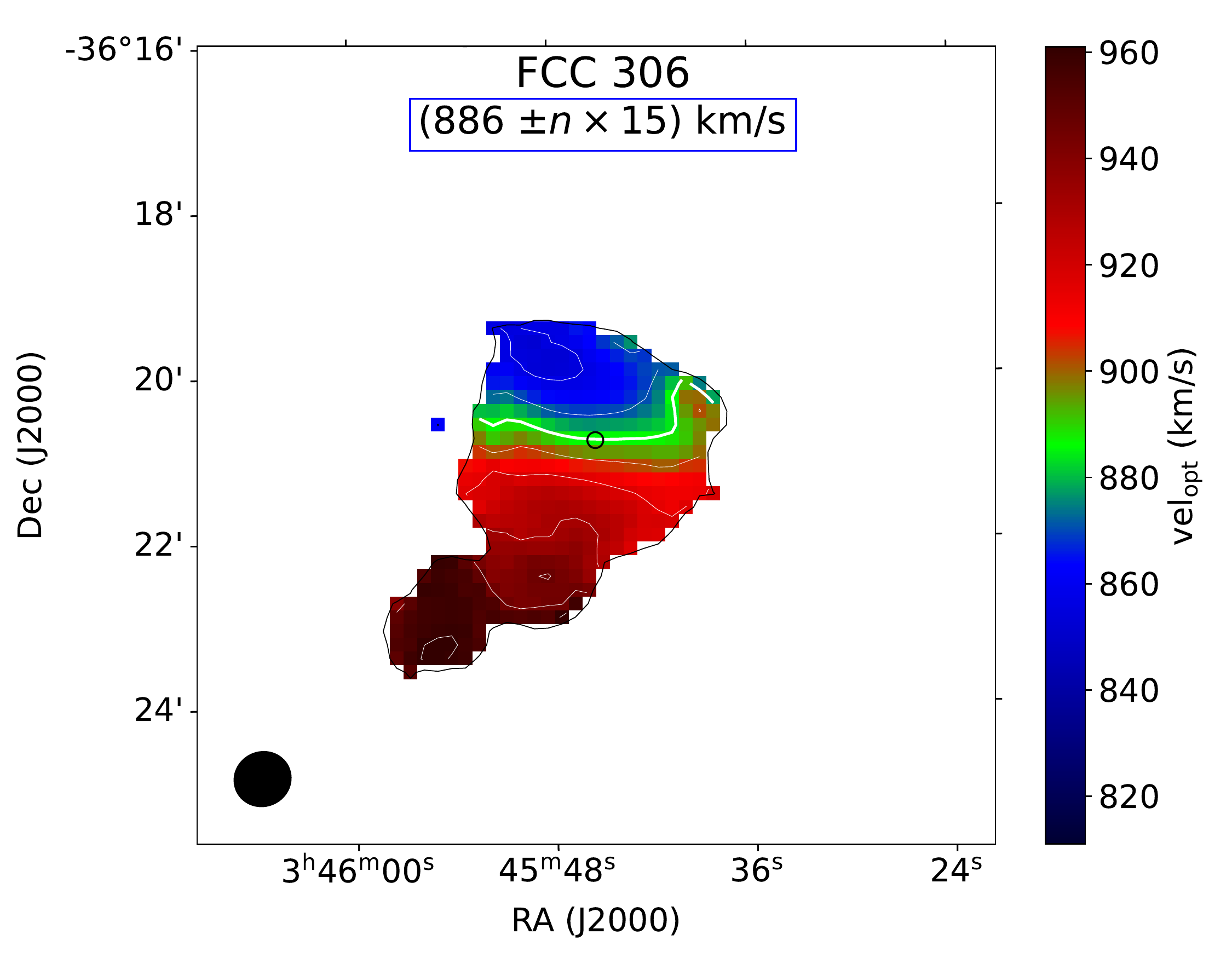}
\includegraphics[width=8.5cm]{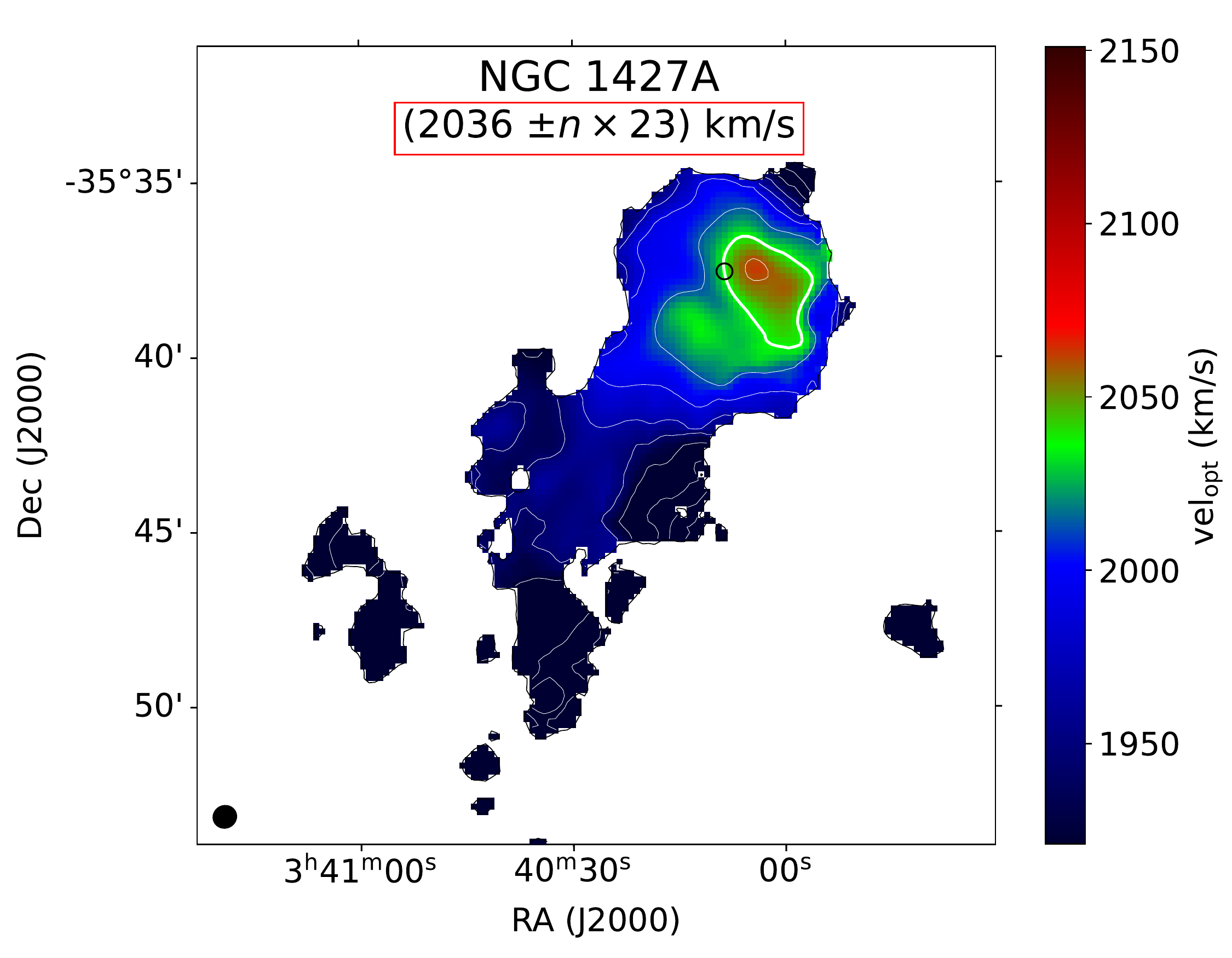}

\includegraphics[width=8.5cm]{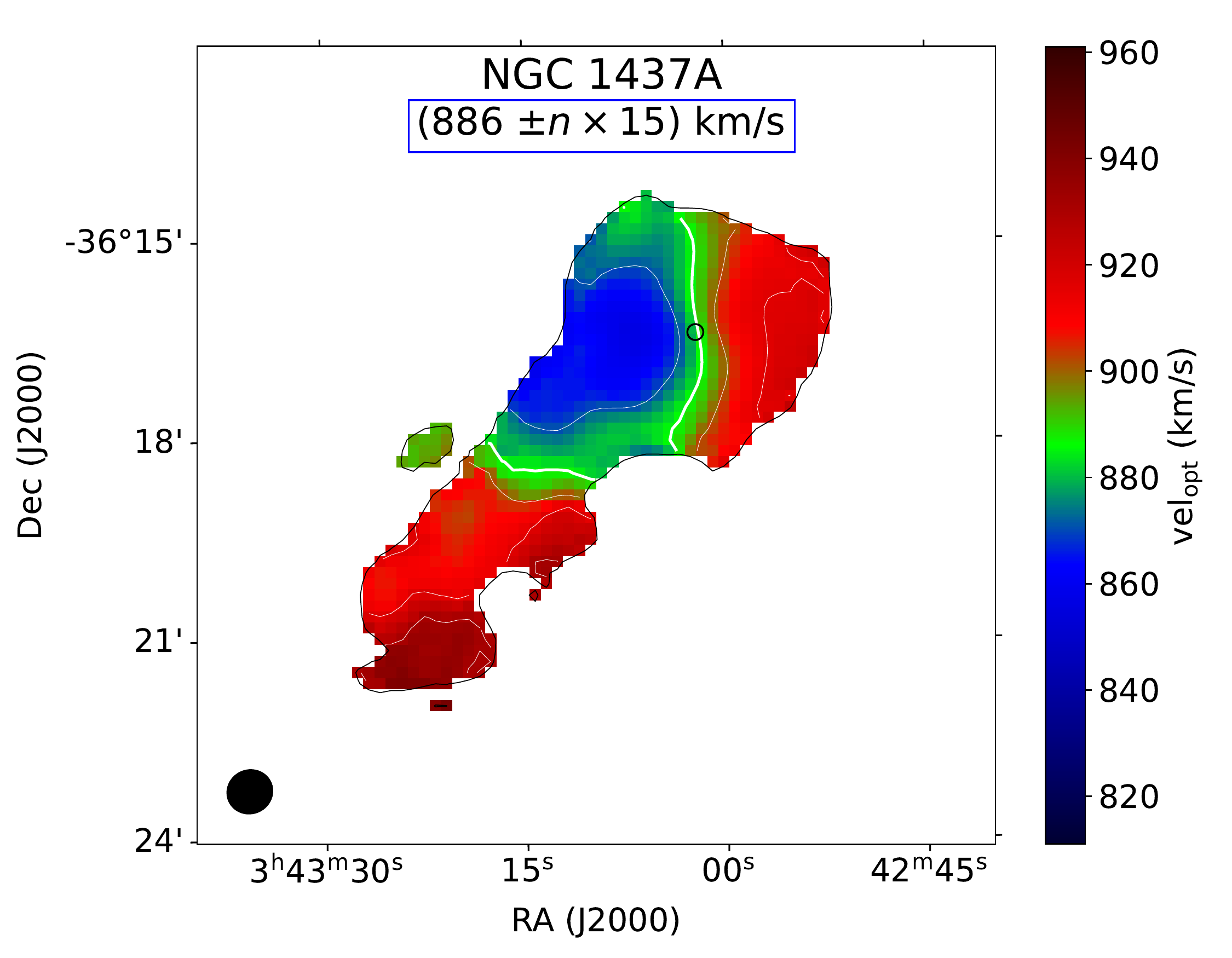}
\includegraphics[width=8.5cm]{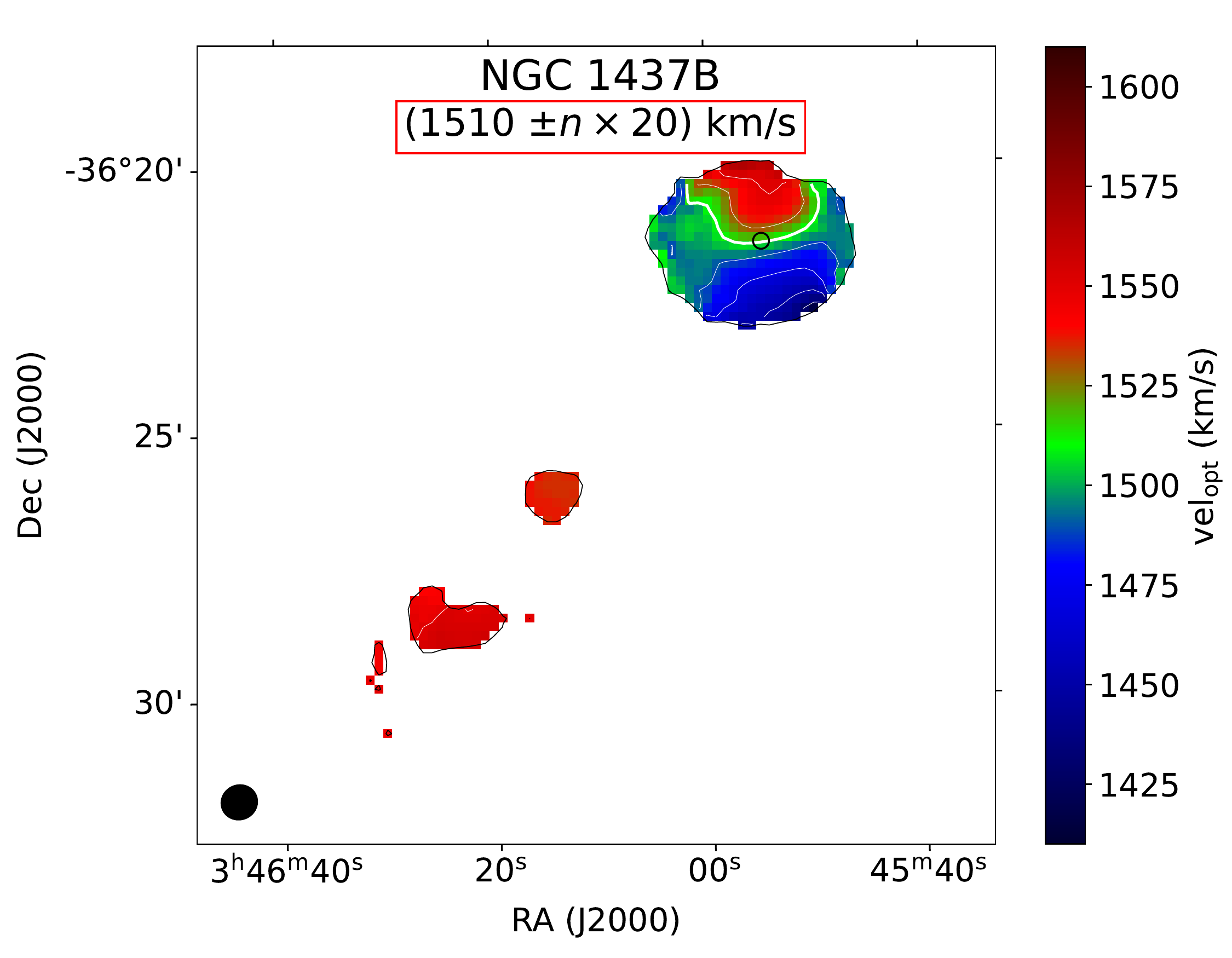}
\caption{Moment-1 \hi\ velocity field of the 6 galaxies with a one-sided \hi\ tail. The galaxy name is given at the top of each image. Below the galaxy name we indicate the \hi\ systemic velocity, which is represented by the thick, white iso-velocity contour in the figure, as well as the levels of the other contours ($n=0, 1, ...$). The coloured frame around the contour levels indicates whether a galaxy is blue- or red-shifted relative the NGC~1399, whose velocity is $\sim1425$ \kms. The bottom-left black ellipse represents the $41''$ \hi\ resolution. The small, open black circle represents the galaxy centre. The thin black contour is the first \hi\ contour shown in Fig. \ref{fig:tails}.}
\label{fig:tailsvf}
\end{figure*}

The \hi\ mosaics of the MeerKAT Fornax Survey improve by 1-2 orders of magnitude over the resolution as well as the \hi\ mass- and column density sensitivity of previous observations of Fornax \citep{kleiner2021,loni2021}. This is bound to deliver new, faint \hi\ detections while at the same time enabling a more complete and detailed study of \hi-bright galaxies already detected in previous observations. In this Section we describe a first key result of our survey related to the latter aspect.

At the time of writing we have covered approximately the eastern half of the survey footprint shown in Fig. \ref{fig:footprint}. Six of our detections stand out for showing long, one-sided \hi\ tails: ESO~358-51, ESO~358-63, FCC~306, NGC~1427A, NGC~1437A and NGC~1437B. All are late-type spirals (from Scd to Irr; \mbox{\citealt{raj2019}}) and all were already known to host \hi. In fact, they represent six of our nine brightest \hi\ detections so far (all with \mhi\ $>10^8$ \msun) --- the other three being NGC~1365, NGC~1436 and ESO~358-60. With the exception of NGC~1427A \citep{leewaddell2018}, none of these six galaxies was previously known to host an \hi\ tail. Even in the case of NGC~1427A, significant new insights can be obtained from the new data. Below we argue that this sample of six galaxies provides the first unambiguous evidence of ram pressure affecting the \hi\ of galaxies in Fornax.

Figure \ref{fig:tails} shows \hi\ contours at an angular resolution of $41''$ ($\sim4$ kpc at the distance of Fornax) overlaid on a $g$-band optical image of the six galaxies taken from the Fornax Deep Survey \citep{iodice2016,venhola2018}. At this resolution, the \hi\ features we identify as one-sided tails are characterised by a low gas density. The most extreme case is that of NGC~1437B, where we only detect two faint clouds south east of the galaxy, making this the only feature whose exact tail nature could be questioned. The only tails reaching a column density above a few times $10^{19}$ cm$^{-2}$ are those of ESO~358-63 and NGC~1427A. In all cases, the tail's length is significantly larger than the size of the stellar body. Most tails are relatively narrow, having a width comparable to that of the $41''$ \hi\ resolution. The clearest exception is that of NGC~1427A, whose tail's width is larger than the size of the host galaxy.

Besides the presence of long one-sided tails, Fig. \ref{fig:tails} shows that the tails do not have a stellar counterpart down to the sensitivity of the optical images. This is typically $\sim28$ mag arcsec$^{-2}$ at full angular resolution (typical seeing $\sim1''$), and fainter when smoothing to lower resolution. In some cases, the sensitivity is worse because of the presence of a nearby bright object, be it a star (as in ESO~358-51) or another galaxy (as in FCC~306). The case of FCC~306 and NGC~1437B deserves an additional note. The two galaxies are very close to one another in projection, and one may suspect that they are interacting. However, our deep optical images do not reveal any clear stellar tails pointing towards such an interaction. Likewise, no \hi\ is detectable in our datacube at the location and at the velocities between the two galaxies. We therefore find no evidence of an on-going interaction. To conclude, the lack of a stellar counterpart of the \hi\ tails shown in Fig. \ref{fig:tails} is a critical consistency argument in favour of ram pressure.

Figure \ref{fig:tails} also shows that all tails are oriented radially within the cluster (the arrow in the top-right corner points towards NGC~1399 at the centre of Fornax). In ESO~358-51 and ESO~358-63 (both in the northern part of Fornax) the tails point towards the cluster centre, while in FCC~306, NGC~1427A, NGC~1437A and NGC~1437B (all in the southern part of Fornax) they point away from it. Considering that \hi-rich galaxies are thought to be recent arrivals in Fornax \citep[e.g.,][]{loni2021} and that most galaxies enter clusters on fairly radial orbits \citep[e.g.,][]{mamon2019}, the tail's orientation is consistent with them being shaped by ram pressure. Furthermore, it suggests that ESO~358-51 and ESO~358-63 are already past pericentre, while FCC~306, NGC~1427A, NGC~1437A and NGC~1437B are still on their way towards it (more on this below).

Fig. \ref{fig:tailsvf} shows the \hi\ velocity fields (same angular resolution as in Fig. \ref{fig:tails}). We find that in galaxies blue-shifted relative to the centre of Fornax (i.e., moving towards us within the cluster and experiencing a red-shifting ram-pressure wind) the velocity of \hi\ in the tail increases with distance from the host galaxy. Conversely, in galaxies red-shifted relative to the centre of Fornax (i.e., moving away from us within the cluster and experiencing a blue-shifting ram-pressure wind) the velocity of \hi\ in the tail decreases as the distance from the host galaxy increases. This result is yet another critical consistency argument in favour of ram pressure shaping the \hi\ tails. The only exception is NGC~1437B. However, this is the galaxy with the least obvious tail and with the smallest velocity offset from the central galaxy NGC~1399 (just 85 \kms), such that it does not represent a striking counter example. Such a small velocity offset should be taken with caution also because what really matters is the motion relative to the local ICM, not to the cluster central galaxy.

In summary, our data reveal a sample of \hi-rich Fornax galaxies with \hi\ tails whose following properties are consistent with ram pressure: \it i) \rm the tails are one sided; \it ii) \rm they are star-less as far as deep optical images allow us to probe; \it iii) \rm they are aligned radially within the cluster; \it iv) \rm the \hi\ velocity gradient along the tails is aligned with the ram-pressure wind given the line-of-sight motion of the host galaxies within the cluster. None of these results alone could be taken as a demonstration that ram pressure is acting on a given galaxy. However, we argue that observing all these properties simultaneously in six of the nine \hi-richest galaxies in our data is an unambiguous indication that ram pressure is indeed at work in Fornax.

The \hi\ tails are thus being shaped by ram pressure, but is ram pressure actually responsible for removing the \hi\ from the stellar body in the first place? An important observation is that all galaxies in this sample exhibit a disturbed stellar body, as discussed in detail by \cite{raj2019}. For example, ESO~358-51 hosts a lopsided stellar disc, which thickens on the side closer to the cluster centre, has an up-bending stellar-light radial profile and shows an isophotal twist (its molecular gas distribution is disturbed, too; \citealt{zabel2019}). Both ESO~358-63 and NGC~1437B have an up-bending radial profile, a boxy, flaring outer disc, an irregular distribution of dust and star forming regions, and outer stellar tails (not coincident with the \hi\ tail). NGC~1427A and NGC~1437A have a strongly lopsided, arrow-shaped stellar body, with a concentration of star-forming regions on the head side of the ``arrow''. Incidentally, this has often been taken as an indication of the direction of motion of these galaxies within the ICM --- towards south west for NGC~1427A and towards south east for NGC~1437A. However, \cite{leewaddell2018} showed that this inference is incorrect in the case of NGC~1427A because it is inconsistent with the direction of the \hi\ tail --- and the same argument can now be applied to NGC~1437A based on the newly detected \hi\ tail. Finally, FCC~306 --- the smallest galaxy in our sample --- exhibits an elongated blue region contained within a redder, circular and slightly offset body.

Based on the literature results summarised above, we conclude that, as a whole, ours is a sample of galaxies that have experienced recent tidal interactions within Fornax. At the moment it is unclear whether these consist of minor mergers, fly-by's or an interaction with the large-scale tidal field of the cluster (in no case can we identify an obvious interacting neighbour, as also discussed by \citealt{leewaddell2018} and \citealt{raj2019}). Regardless, the connection between a tidally disturbed stellar body and the presence of an \hi\ tail is clear. Of the other relatively bright \hi\ detections observed so far with MeerKAT and mentioned above, NGC~1436 and ESO~358-60 show no indication of recent tidal interactions and no \hi\ tail\footnote{We do not discuss NGC~1365 here as it is a unique object in the context of the \hi-rich galaxy population of Fornax. This galaxy has a much higher \mstar\ and \mhi\ than all other \hi-detected galaxies together. Its complex and very extended \hi\ disc will be analysed in detail in a separate paper.}. This connection leads us to argue that the \hi\ currently in the tails may not have been stripped by ram pressure. Rather, tidal interactions may be responsible for pulling some \hi\ out of the stellar body. Once there, the cold gas is more susceptible to being displaced further by the weak ram pressure expected in a small cluster like Fornax. This mechanism was proposed to explain some of the \hi\ tails in Virgo \citep{vollmer2003,chung2007,chung2009}, and should be even more relevant in Fornax, where $\rho_\mathrm{ICM} \cdot v_\mathrm{gal}^2$ is about an order of magnitude lower. In this picture, ram pressure does not strip \hi\ from galaxies in Fornax, but it shapes tidally-stripped \hi\ into the one-sided, radially-oriented, star-less tails we observe.

The tidal-then-ram pressure mixed origin of the \hi\ tails in Fornax is consistent with some tails appearing past pericenter and, therefore, pointing towards the cluster centre (ESO~358-51 and ESO~358-63). The reason might be that only at that point of their orbit the host galaxies experienced the tidal interaction required to move some \hi\ to larger radius and make it susceptible to further displacement through ram pressure.

\section{Summary}
\label{sec:summary}

We have presented the MeerKAT Fornax Survey --- a deep, high-resolution MeerKAT observation of the Fornax galaxy cluster performed with the aim of advancing our understanding of galaxy evolution in low-redshift, low-mass clusters. We have described the survey design and observations, and we have presented a detailed discussion of the \hi\ data processing to provide a reference for future papers. The 12 deg$^2$ survey footprint covers the cluster central region out to $\sim R_\mathrm{vir}$ and extends to $\sim2 R_\mathrm{e}$ towards south west to include the NGC~1316 galaxy group. The \hi\ mosaic cubes have an angular resolution from $\sim10''$ to $\sim100''$, a velocity resolution of 1.4 \kms, a column density sensitivity between $\sim8\times10^{17}$ and $5\times10^{19}$ cm$^{-2}$ depending on resolution, and an \hi\ mass sensitivity of $\sim6\times10^5$ \msun.

We have presented a first key result from the data obtained so far. We find that six of the nine brightest \hi\ detections so far have a long, one-sided, star-less \hi\ tail aligned radially within the cluster. Only one of these tails was previously known. The \hi\ velocity gradient along the tails is aligned with the direction of the ram-pressure wind given the motion of the host galaxies within the ICM. All these properties are consistent with ram pressure shaping the tails. We thus argue that this sample of \hi\ tails represents the first unambiguous observational evidence of ram pressure affecting the evolution of galaxies falling into Fornax.

All six galaxies hosting an \hi\ tail exhibit signatures of recent tidal interactions, unlike other Fornax \hi-bright galaxies without a tail. We therefore argue that there is a connection between tidal interactions and the formation of ram-pressure \hi\ tails in Fornax. Considering that ram pressure is expected to be relatively weak in Fornax (e.g., $\sim$ one order of magnitude weaker than in Virgo), we argue that \hi\ in the tails was pulled out of the stellar body by tidal forces and, once there, was more susceptible to being further displaced by ram pressure. That is, ram pressure did not remove the \hi\ from the stellar body, but shaped \hi\ stripped through tidal forces into a star-less tail with the direction, length and velocity gradient we have now observed with MeerKAT.

More lessons on galaxy evolution in Fornax are likely to be learnt from the analysis of the \hi\ tails presented in this paper (as well as all of other \hi\ detections) using the high-resolution data products delivered by the MeerKAT Fornax Survey (Table \ref{table:mos}). This will be the subject of future work.

\begin{acknowledgements}

This article is dedicated to the memory of our fellow MeerKAT Fornax Survey team member Sergio Colafrancesco. We are grateful to the full MeerKAT team for their work building, commissioning and operating MeerKAT, and for their support to the MeerKAT Fornax Survey. We acknowledge the outstanding support given by the Ilifu staff, in particular Jordan Collier and Jeremy Smith. The MeerKAT telescope is operated by the South African Radio Astronomy Observatory, which is a facility of the National Research Foundation, an agency of the Department of Science and Innovation. We acknowledge the use of the Ilifu cloud computing facility - www.ilifu.ac.za, a partnership between the University of Cape Town, the University of the Western Cape, the University of Stellenbosch, Sol Plaatje University, the Cape Peninsula University of Technology and the South African Radio Astronomy Observatory. The Ilifu facility is supported by contributions from the Inter-University Institute for Data Intensive Astronomy (IDIA - a partnership between the University of Cape Town, the University of Pretoria and the University of the Western Cape), the Computational Biology division at UCT and the Data Intensive Research Initiative of South Africa (DIRISA). This project has received funding from the European Research Council (ERC) under the European Union's Horizon 2020 research and innovation programme (grant agreement no. 679627, ``FORNAX''; and grant agreement no. 882793, ``MeerGas''). The data of the MeerKAT Fornax Survey are reduced using the CARACal pipeline, partially supported by ERC Starting grant number 679627, MAECI Grant Number ZA18GR02, DST-NRF Grant Number 113121 as part of the ISARP Joint Research Scheme, and BMBF project 05A17PC2 for D-MeerKAT. Information about CARACal can be obtained online under the URL: https://caracal.readthedocs.io. At RUB, this research is supported by the BMBF project 05A20PC4 for D-MeerKAT. We acknowledge the support from the Ministero degli Affari Esteri della Cooperazione Internazionale - Direzione Generale per la Promozione del Sistema Paese Progetto di Grande Rilevanza ZA18GR02. This work is based on research supported by the National Research Foundation of South Africa (Grant Number 113121). FL acknowledges financial support from the Italian Ministry of University and Research - Project Proposal CIR01-00010. GLB acknowledges support from the NSF (AST-2108470, ACCESS MCA06N030), NASA TCAN award 80NSSC21K1053, and the Simons Foundation. DJP is supported through the South African Research Chairs Initiative of the Department of Science and Technology and National Research Foundation.

\end{acknowledgements}

%
%

\bibliographystyle{aa} 
\bibliography{myrefs} 

\begin{appendix}

\section{MeerKAT Fornax Survey pointings}

 \longtab[1]{
 \begin{longtable}{lll}
\caption{\label{table:radec} RA and Dec of the 91 telescope pointings of the MeerKAT Fornax Survey. With reference to Fig. \ref{fig:footprint}, the pointings are ordered from east to west and, at approximately fixed RA, from north to south.}\\
\hline\hline
filed & RA$_\mathrm{J2000}$ & Dec$_\mathrm{J2000}$ \\
        &  (h:m:s) & (d:m:s) \\
\hline
\endfirsthead
 \caption{continued.}\\
\hline\hline
field & RA$_\mathrm{J2000}$ & Dec$_\mathrm{J2000}$ \\
        &  (h:m:s) & (d:m:s) \\
\hline
\endhead
\hline
\endfoot
		01 & 03:49:08.3 & -35:56:57 \\
		02 & 03:49:12.0 & -36:23:54 \\									
		03 & 03:47:05.7 & -34:49:33 \\
		04 & 03:47:08.5 & -35:16:31 \\
		05 & 03:47:11.4 & -35:43:28 \\
		06 & 03:47:14.4 & -36:10:25 \\
		07 & 03:47:17.5 & -36:37:23 \\
		08 & 03:45:10.8 & -34:36:05 \\
		09 & 03:45:13.0 & -35:03:02 \\
		10 & 03:45:15.3 & -35:29:59 \\
		11 & 03:45:17.6 & -35:56:57 \\
		12 & 03:45:19.9 & -36:23:54 \\
		13 & 03:45:22.3 & -36:50:52 \\
		14 & 03:43:16.6 & -34:22:36 \\
		15 & 03:43:18.2 & -34:49:33 \\
		16 & 03:43:19.8 & -35:16:31 \\
		17 & 03:43:21.4 & -35:43:28 \\
		18 & 03:43:23.1 & -36:10:25 \\
		19 & 03:43:24.8 & -36:37:23 \\
		20 & 03:43:26.5 & -37:04:20 \\
		21 & 03:41:23.0 & -34:09:07 \\
		22 & 03:41:23.9 & -34:36:05 \\
		23 & 03:41:24.9 & -35:03:02 \\
		24 & 03:41:25.9 & -35:29:59 \\
		25 & 03:41:26.9 & -35:56:57 \\
		26 & 03:41:27.9 & -36:23:54 \\
		27 & 03:41:28.9 & -36:50:52 \\
		28 & 03:39:30.0 & -33:55:39 \\
		29 & 03:39:30.3 & -34:22:36 \\
		30 & 03:39:30.7 & -34:49:33 \\
		31 & 03:39:31.0 & -35:16:31 \\
		32 & 03:39:31.4 & -35:43:28 \\
		33 & 03:39:31.7 & -36:10:25 \\
		34 & 03:39:32.1 & -36:37:23 \\
		35 & 03:37:37.3 & -34:09:07 \\
		36 & 03:37:37.1 & -34:36:05 \\
		37 & 03:37:36.8 & -35:03:02 \\
		38 & 03:37:36.5 & -35:29:59 \\
		39 & 03:37:36.2 & -35:56:57 \\
		40 & 03:37:35.9 & -36:23:54 \\
		41 & 03:35:44.1 & -34:22:36 \\
		42 & 03:35:43.2 & -34:49:33 \\
		43 & 03:35:42.2 & -35:16:31 \\
		44 & 03:35:41.3 & -35:43:28 \\
		45 & 03:35:40.3 & -36:10:25 \\
		46 & 03:35:39.4 & -36:37:23 \\
		47 & 03:33:50.2 & -34:36:05 \\
		48 & 03:33:48.6 & -35:03:02 \\
		49 & 03:33:47.1 & -35:29:59 \\
		50 & 03:33:45.5 & -35:56:57 \\
		51 & 03:33:43.8 & -36:23:54 \\
		52 & 03:33:42.2 & -36:50:52 \\
		53 & 03:31:55.6 & -34:49:33 \\
		54 & 03:31:53.5 & -35:16:31 \\
		55 & 03:31:51.3 & -35:43:28 \\
		56 & 03:31:49.0 & -36:10:25 \\
		57 & 03:31:46.7 & -36:37:23 \\
		58 & 03:31:44.3 & -37:04:20 \\
		59 & 03:30:00.5 & -35:03:02 \\
		60 & 03:29:57.7 & -35:29:59 \\
		61 & 03:29:54.8 & -35:56:57 \\
		62 & 03:29:51.8 & -36:23:54 \\
		63 & 03:29:48.8 & -36:50:52 \\
		64 & 03:29:45.7 & -37:17:49 \\
		65 & 03:28:04.7 & -35:16:31 \\
		66 & 03:28:01.2 & -35:43:28 \\
		67 & 03:27:57.6 & -36:10:25 \\
		68 & 03:27:54.0 & -36:37:23 \\
		69 & 03:27:50.2 & -37:04:20 \\
		70 & 03:27:46.4 & -37:31:18 \\
		71 & 03:26:08.3 & -35:29:59 \\
		72 & 03:26:04.1 & -35:56:57 \\
		73 & 03:25:59.8 & -36:23:54 \\
		74 & 03:25:55.4 & -36:50:52 \\
		75 & 03:25:50.9 & -37:17:49 \\
		76 & 03:25:46.4 & -37:44:46 \\
		77 & 03:24:11.2 & -35:43:28 \\
		78 & 03:24:06.3 & -36:10:25 \\
		79 & 03:24:01.3 & -36:37:23 \\
		80 & 03:23:56.2 & -37:04:20 \\
		81 & 03:23:50.9 & -37:31:18 \\
		82 & 03:23:45.6 & -37:58:15 \\
		83 & 03:22:13.4 & -35:56:57 \\
		84 & 03:22:07.8 & -36:23:54 \\
		85 & 03:22:02.0 & -36:50:52 \\
		86 & 03:21:56.2 & -37:17:49 \\
		87 & 03:21:50.2 & -37:44:46 \\
		88 & 03:20:14.9 & -36:10:25 \\
		89 & 03:20:08.6 & -36:37:23 \\
		90 & 03:20:02.1 & -37:04:20 \\
		91 & 03:19:55.5 & -37:31:18 \\
\end{longtable}
}

\end{appendix}

\end{document}